\documentclass[12pt]{article}

\usepackage{enumerate}

\usepackage{url} % not crucial - just used below for the URL 

\newcommand{\blind}{1}

\usepackage[english]{babel}
\usepackage[utf8x]{inputenc}
\usepackage[T1]{fontenc}
\usepackage{natbib}
\usepackage{cancel}
\usepackage{soul}
 \bibpunct{(}{)}{;}{a}{,}{,}

\usepackage{fullpage}

\usepackage{amsmath,amsthm, amssymb}
\usepackage{bm}
\usepackage{bbm} %% indicator function
\usepackage{graphicx}
\usepackage[colorinlistoftodos]{todonotes}
\usepackage[colorlinks=true, allcolors=blue]{hyperref}
\usepackage{enumitem}
\usepackage{algpseudocode}
\usepackage{multirow}
\usepackage{comment}

\usepackage{setspace}
\usepackage{float} %float figure
\usepackage{caption}
\usepackage{subcaption}
\usepackage[export]{adjustbox} %adjust vertical position of graphs 

\usepackage{booktabs}
\usepackage{longtable}
\usepackage{array}
\usepackage{multirow}
\usepackage{wrapfig}
\usepackage{float}
\usepackage{colortbl}
\usepackage{pdflscape}
\usepackage{tabu}
\usepackage{threeparttable}
\usepackage{threeparttablex}
\usepackage[normalem]{ulem}
\usepackage{makecell}
\usepackage{xcolor}
\newtheorem{theorem}{Theorem}
\newtheorem{fact}{Fact}
\newtheorem{corollary}{Corollary}[theorem]

\newtheorem{proposition}{Proposition}
\theoremstyle{definition}
\newtheorem{definition}{Definition}
\newtheorem{assumption}{Assumption}

\usepackage[linesnumbered,ruled,vlined]{algorithm2e}

\usepackage{scalerel,stackengine}
\stackMath
\newcommand\reallywidehat[1]{%
\savestack{\tmpbox}{\stretchto{%
  \scaleto{%
    \scalerel*[\widthof{\ensuremath{#1}}]{\kern-.6pt\bigwedge\kern-.6pt}%
    {\rule[-\textheight/2]{1ex}{\textheight}}%WIDTH-LIMITED BIG WEDGE
  }{\textheight}% 
}{0.5ex}}%
\stackon[1pt]{#1}{\tmpbox}%
}
\newcommand{\te}{\tau}
\newcommand{\teNaive}{\check{\tau}}

\newcommand{\teNaiveNull}{\check{\tau}_{\mathcal{C}}}

\newcommand{\rsqtx}{R^2_{T \sim U \mid X}}
\newcommand{\rsqt}{R^2_{T \sim U}}
\newcommand{\rsqytx}{R^2_{a'Y \sim U \mid T,X}}
\newcommand{\muU}{\mu_{u\mid t,x}}
\newcommand{\SigmaU}{\Sigma}
\newcommand{\sigmat}{\sigma}

\usepackage{graphicx,scalerel}

\begin{document}

\if1\blind
{
  \title{\bf Sensitivity to Unobserved Confounding in Studies with Factor-structured Outcomes}

  \author{Jiajing Zheng\\UCSB \and  Jiaxi Wu\\UCSB \and Alexander D'Amour \\Google Research \and Alexander Franks\\UCSB}
\date{\today}
  
%   \author{Author 1\thanks{
%     The authors gratefully acknowledge \textit{please remember to list all relevant funding sources in the unblinded version}}\hspace{.2cm}\\
%     Department of YYY, University of XXX\\
%     and \\
%     Author 2 \\
%     Department of ZZZ, University of WWW}
  \maketitle
} \fi

\if0\blind
{
  \bigskip
  \bigskip
  \bigskip
  \begin{center}
    {\large\bf Sensitivity to Unobserved Confounding in Studies with Factor-structured Outcomes}
\end{center}
  \medskip
} \fi

\def\-{\text{-}}
\def\Bern{\text{Bern}}

\def\PATE{$\text{PATE}_{\Delta t}$ }
\def\Bias{$\text{Bias}_{\Delta t}$ }

\newcommand\sbullet[1][.5]{\mathbin{\ThisStyle{\vcenter{\hbox{%
  \scalebox{#1}{$\SavedStyle\bullet$}}}}}%
}

%% For editing
\newcommand\cmnt[2]{\;
{\textcolor{red}{[{\em #1 --- #2}] \;}
}}
\newcommand\jiajing[1]{\cmnt{#1}{Jiajing}}
\newcommand\damour[1]{\cmnt{#1}{D'Amour}}
\newcommand\franks[1]{\cmnt{#1}{AF}}

\thispagestyle{empty}
\pagenumbering{gobble}

% \newpage
\begin{abstract}

In this work, we propose an approach for assessing sensitivity to unobserved confounding in studies with multiple outcomes. We demonstrate how prior knowledge unique to the multi-outcome setting can be leveraged to strengthen causal conclusions beyond what can be achieved from analyzing individual outcomes in isolation.  We argue that it is often reasonable to make a shared confounding assumption, under which residual dependence amongst outcomes can be used to simplify and sharpen sensitivity analyses.  We focus on a class of factor models for which we can bound the causal effects for all outcomes conditional on a single sensitivity parameter that represents the fraction of treatment variance explained by unobserved confounders.  We characterize how causal ignorance regions shrink under additional prior assumptions about the presence of null control outcomes,  and provide new approaches for quantifying the robustness of causal effect estimates.  Finally, we illustrate our  sensitivity analysis workflow in practice, in an analysis of both simulated data and a case study with data from the National Health and Nutrition Examination Survey (NHANES).

\vspace{2em}
\end{abstract}

%%%%%%%%% INTRODUCTION %%%%%%%%%%%%%%%%%%
\clearpage
\pagenumbering{arabic}

\doublespacing

\section{Introduction}

Large observational datasets often include measurements of multiple outcomes of interest. For example, in high-throughput biology, the goal might be to understand the effect of a treatment on multiple biomarkers \citep{leek2007capturing, zhao2018cross}, or in patient-centered epidemiologic studies, researchers might study the potential impact of health recommendations on multiple disease-related outcomes \citep{sanchez2005structural, kennedy2019estimating}. While it is always possible to analyze each outcome separately, there has been a recent emphasis on the importance of techniques for simultaneously inferring the effect of an intervention on multiple outcomes \citep[][]{vanderweele2017outcome}.

As in any observational study, the validity of multi-outcome causal inference rests on untestable and often implausible assumptions about unconfoundedness.  As such, methods which explore the sensitivity and robustness of effects to assumptions about unconfoundedness are increasingly recognized as a crucial part of any rigorous analysis.  In this paper, we demonstrate how to leverage prior knowledge to strengthen causal conclusions from observational datasets with multiple outcomes. We propose a sensitivity analysis unique to multi-outcome inference, which leads to characterizations of the robustness of causal effects that are both simpler and sharper than those that can be achieved by analyzing each outcome separately. We focus on linear factor models of the outcomes where we 1) establish bounds on the magnitude of unobserved confounding bias for all outcomes as a function of a single interpretable sensitivity parameter under an assumption about shared confounding,  2) provide novel theoretical results that demonstrate how assumptions about null control outcomes inform the sensitivity analysis and 3) provide practical guidance and a workflow for sensitivity analysis for multi-outcome studies. We demonstrate this workflow in simulation and an analysis of the effect of light drinking on multiple biomarkers for health. 

After reviewing related literature in Section \ref{sec:related_lit}, we introduce the problem setting and the challenges in multi-outcome causal inference (Section \ref{sec:copula-based_sens_analy_framework, multi-y}). In Section \ref{sec:copula-based_sens_analy_gaussian_copula,multi-y}, we establish the framework for our proposed sensitivity analysis with multivariate outcomes and introduce theoretical results about confounding bias for a model in which the expected outcomes are linear in the unobserved confounders (given observed covariates). We provide additional novel theories demonstrating how null control outcomes can be leveraged in conjunction with our sensitivity analysis and how to characterize the robustness of treatment effect estimates. We illustrate the theoretical insights with a Bayesian implementation of our sensitivity analysis on simulated data in Section \ref{sec:simulation}. In Section \ref{sec:calibration,multi-y}, we discuss the interpretation and calibration of sensitivity parameters. In Section \ref{sec:nhanes}, we illustrate our approach on a real-world example of the effects of light drinking on health measures using data from the National Health and Nutrition Examination Study (NHANES).

% Building on \citet{wang2017confounder} and \citet{gagnon2012using}, w

\subsection{Related Literature}
\label{sec:related_lit}
There are a number of methods for causal inference with multiple outcomes, although the majority of these appear in the context of randomized control trials \citep[][]{freemantle2003composite, mattei2013exploiting}.  In the context of observational studies, \citet{sammel1999multivariate} propose a multivariate linear mixed effects model for multi-outcome causal inference, \citet{thurston2009bayesian} consider a Bayesian generalization of multiple outcomes which are nested across different domains, and \citet{sanchez2005structural} review a variety of structural equation models for multiple outcomes with application to epidemiological problems. These works have the important caveat that they typically assume a version of the ``no unobserved confounding'' (NUC) assumption.  More recently, \citet{kennedy2019estimating} develop a nonparametric doubly robust method for estimation and hypothesis testing of scaled treatment effects on multiple outcomes, but again assume NUC.

When unobserved confounding is expected to be an issue, certain assumptions about additional outcomes can sometimes be used to identify the effects of alternative primary outcomes.  As one important example, additional outcomes called ``proxy confounders'' or ``null control outcomes'' can sometimes be used to identify causal effects for a set of primary outcomes \citep{shi2020selective, tchetgen2020introduction}.  Relatedly, \citet{wang2017confounder} establish identification assumptions in a linear Gaussian model with multiple outcomes when the non-null effects are assumed to be sparse.  Although not explicitly framed in causal language, a closely related line of work explores approaches for de-biasing estimates in the presence of confounders, for example when there are batch effects in high-throughput biological datasets \citep[e.g.][]{gagnon2012using, gagnon2013removing}.  These works all focus on various assumptions that can be made to point identify causal effects for each outcome.

In contrast, sensitivity analyses are useful for explicitly relaxing such identification assumptions. There are a variety of different approaches for assessing sensitivity to potential unmeasured confounders in single-treatment, single-outcome settings \citep{rosenbaum1983assessing, tan2006distributional, franks2019flexible, cinelli2020making, veitch2020sense}. In the multi-outcome setting, \citet{fogarty2016sensitivity}
consider sensitivity analysis for multiple comparisons in matched observational studies using weighting estimators. They focus on the implications of the key fact that omitted variable biases across multiple outcomes are connected through the shared effect of the unmeasured confounders on the treatment.  Under a similar matched pairs framework, \citet{rosenbaum2021sensitivity} considers a sensitivity analysis for a single primary outcome and shows how a null control outcome can sometimes increase the evidence for the robustness of the primary outcome.  One potential concern with their sensitivity analyses for weighting estimators is that the sensitivity analyses are implicitly based on the overly conservative assumption that unmeasured confounders explain nearly all the variation in the outcomes. We address this concern by developing a sensitivity analysis in which we explicitly account for the effect of unmeasured confounders on the outcomes.

% In our work, we provide similar insights about the role of null control outcomes, although from a much different perspective. 

Building on existing parametric models for multiple outcomes \citep[e.g.][]{sammel1999multivariate, sanchez2005structural, thurston2009bayesian}, we propose an outcome model with latent variables which account for the residual correlations between outcomes. Unlike these previous works, we expressly consider the possibility that these latent variables might correspond to potential confounders, in that they may also correlate with the treatment.  To account for the potential dependence between latent variables and treatment, we use a so-called latent variable sensitivity analysis \citep{rosenbaum1983assessing}.  Typically, in a latent variable sensitivity analysis, sensitivity parameters govern the functional relationship between a latent variable and both the outcome and the treatment.  For instance, \citet{cinelli2020making} propose a particularly intuitive latent variable sensitivity analysis for single-treatment, single-outcome problems, with two sensitivity parameters corresponding to the fraction of outcome residual variance explained by latent confounders and the fraction of treatment residual variance explained by confounders.  We generalize their approach to multi-outcome settings, paralleling a sensitivity analysis strategy developed in the context of causal inference with multiple treatments and a scalar outcome \citep{zheng2021copula}.  

% We also implement a Bayesian approach for inference and partial identification, similar to \citet{zheng2021bayesian}, who emphasize the importance of prior specifications for multiple partially identified treatment effects.

% More studies for causal inference problems with multivariate outcomes can be found in \citep{jo2001modeling,hernan2002estimating,flanders2015general,li2017general}.  

\section{Setup}
\label{sec:copula-based_sens_analy_framework, multi-y}

We let $Y$ denote a q-vector of outcomes, $T$ a scalar treatment variable, $U$ an $m$-vector of potential unobserved confounders, and $X$ any observed pre-treatment covariates. The goal of multi-outcome causal inference is to infer the effect of a scalar treatment on the q-dimensional outcomes.  In this setting, we define a class of causal estimands as the population average treatment effect (PATE) for any linear combination of the outcomes, $a'Y$, as
\begin{equation}
    \text{PATE}_{a,t_1, t_2} := E[a'Y \mid do(t_1)] - E[a'Y \mid do(t_2)],
\end{equation}
\noindent where the \emph{do}-operator indicates that we are intervening to set the level to $t$ rather than merely conditioning on $t$  \citep{pearl2009causality}. Most commonly, we take $a = e_j$ for some $j$, so that $PATE_{e_j, t_1, t_2}$ is simply the causal effect on measured outcome $j$ in the original coordinates. In some cases, other linear combinations may be of interest, for example when there is not enough power to detect differences in individual outcomes, but there are detectable and interesting differences for linear combinations of outcomes \citep[e.g. see][]{cook2010envelope}.  Relatedly, it is often desirable to define the PATE on standardized outcomes, so that each dimension of $Y$ has unit variance e.g. $a_j = e_j/sd(Y_j)$ \citep{kennedy2019estimating}.

In general, the PATEs cannot be identified from observational data without assumptions about the absence of unobserved confounders.  If, in addition to $X$, the unmeasured confounders $U$ were observed, then the following three assumptions would be sufficient to identify the causal effects: 

\begin{assumption}[Latent unconfoundedness]
\label{asm:latent-unconfoundedness-scm}
$U$ and $X$ block all backdoor paths between $T$ and $Y$ \citep{pearl2009causality}.
\end{assumption}
\begin{assumption}[Latent positivity]
\label{asm:latent-positivity-scm}
$f(T = t \mid U = u, X=x) > 0$ for all $u$ and $x$.
\end{assumption}
\begin{assumption}[SUTVA]\label{asm:sutva} There is no hidden version of the treatment and no interference between units \citep[][]{rubin1980comment}.
\end{assumption}

\noindent  Since $U$ is not observed, we cannot identify causal effects without additional assumptions. Instead of making potentially implausible assumptions about NUC, we advocate for reasoning about the strength of potential unobserved confounding.  Here, we argue that sensitivity analysis can be a useful tool for characterizing how robust our multiple causal conclusions are to such confounding.

% Same as the multi-treatment case, the multi-outcome settings introduce new observable implications, which makes identifiability issue more complicated, but, in the meantime, could be useful for providing additional information about confounders.
% In the multiple outcome settings, unlike the multi-treatment cases, the conditional confounder distribution $f_{\psi_T}(u \mid t)$ is not identifiable, but, instead, the conditional outcome distribution $f_{\psi_Y}(y \mid t, u)$ can be partially identified from the multivariate correlation structure of the outcomes. We focus on this specific case where the relationship between confounders and outcomes are partially identifiable  under appropriate assumptions. 
% Thus, like the multiple treatment settings, to execute decent sensitivity analyses in the multi-outcomes cases, sensitivity analysis parameters must not only be uninformed by the observed data distribution, but the parameter describing the outcome-confounder relationship must also be decoupled from the parameter describing the treatment-confounder relationship. As shown in Section \ref{sec:latent-confounder}, these can be achieved through our copula-based factorization. 

% In this section, we discuss how copula-based factorization can be applied to the multiple outcome setting, and elaborate its implementation with the Gaussian copula in particular.

\section{Sensitivity Analysis with Multiple Outcomes}
\label{sec:copula-based_sens_analy_gaussian_copula,multi-y}

Let $f(y \mid do(t))$ denote the distribution of $y$ if we were to intervene to set the level of treatment to $t$. As shown in  \citet{damour2019aistats} and \citet{zheng2021copula}, $f(y \mid do(t))$ can be written as $f_{\psi}(y \mid do(t)) = \int_{\mathcal{U}} f_{\psi_{Y}}(y \mid t, u)\left[\int f_{\psi_{T}}(u \mid \tilde{t}) f(\tilde{t}) d \tilde{t} \right] d u$ where $f_{\psi_{Y}}(y \mid t, u)$ is the full conditional outcome density, $f_{\psi_{T}}(u \mid t)$ is a proposed distribution for unobserved confounders given the treatment and $f(t)$ is the marginal density of the treatment.  In the multivariate outcome setting, $\psi_Y$ are sensitivity parameters which govern the relationship between the $q$-dimensional outcomes and the $m$-dimensional unobserved confounders, whereas $\psi_T$ are parameters which govern the relationship between the scalar treatment and unobserved confounders.  

A potential difficulty with multi-outcome sensitivity analysis is that the dimension of $\psi_Y$ scales with the number of outcomes. However, with multiple outcomes, there is also often additional prior knowledge that can be brought to bear on the problem.  We consider two such assumptions that mitigate the challenges associated with having to reason about high-dimensional sensitivity parameters, and can, in some cases, tighten bounds beyond what we would obtain from analyzing outcomes one at a time.  First, we explore the implications of assuming that confounding is shared across multiple outcomes.  Second, we consider how the sensitivity analysis changes under the additional assumption that there is no causal effect for some specific outcomes (null controls).

To begin, in Section \ref{sec:shared_confounding} we propose our model for multi-outcome causal inference and then in Section \ref{sec:bias_nonGaussian}, we establish our sensitivity parameterization and derive worst-case bounds on the causal effects for all outcomes.  For this model, under standard identifiability conditions for factor models, we show that the relevant dimensions of $\psi_Y$ are identified and the bound on the magnitude of the confounding bias for all outcomes depends on a single unknown scalar sensitivity parameter governing the strength of the confounder-treatment relationship.  In Section \ref{sec:nco}, we characterize the joint relationship between the confounding biases across outcomes by establishing how assumptions about null control outcomes constrain the set of plausible causal conclusions. Finally, we discuss different robustness measures in Section \ref{sec:cal_rv}.

% We demonstrate the behavior of the sensitivity analysis and highlight the theoretical implications from this section on simulated data in Section \ref{sec:simulation}.

% In Appendix \ref{sec:relaxing}, we discuss relaxations of our proposed shared confounding assumption.

% \begin{definition}[Causal equivalence class for multi-outcome inference]
% Let $\tilde u = h(u)$ where $h$ is any bijective function of the unobserved confounder.  Then, $[\psi_Y]$ is a causal equivalence class of $\psi_Y$ if and only if for any $\tilde \psi_Y$ in $[\psi_Y]$, then, for every $\psi_T$ there exists a $\tilde \psi_{T}$ and $h$ with $f_{\psi_T}(t \mid u) = f_{\tilde \psi_T}(t \mid \tilde u)$ such that $f_{\psi_Y, \psi_T}(y \mid do(T = t)) = f_{\tilde \psi_{Y}, \tilde \psi_{T}}(y \mid do(T = t))$ for all $y, t$.
% \end{definition}

\subsection{A Multi-outcome Model with Factor Confounding}
\label{sec:shared_confounding}

In this paper, we focus on models for observed outcomes which have factor-structured residuals. We seek to leverage the factor-structured residuals for assessing the sensitivity of causal conclusions to assumptions about unobserved confounders.  We define the conditional mean and covariance of the observed outcome distribution to be
\begin{equation}
    E[Y \mid T = t, X=x] := \check g(t, x) \text{ and } Cov(Y \mid T = t, X = x) = \Gamma_{t,x}\Gamma_{t,x}' + \Delta, \label{eqn:obs_moments}
\end{equation}
\noindent where $\Gamma_{t, x}$ are rank-$m$ factor loadings and $\Delta$ is a constant diagonal matrix.  For now, we assume the factor loadings can vary with both $t$ and $x$, but later make the stronger assumption that the factor loadings are constant.

There are several causal models consistent with the observed data moments in \eqref{eqn:obs_moments}.  Here, we propose a causal model which is explicitly parameterized in terms of latent factors, $U$, motivated by the idea that unmeasured confounders can induce residual correlation between outcomes.  Throughout, we assume the following structural equation model:
% \begin{assumption}[Conditionally Gaussian outcomes and confounders.]
% \label{asm:copula}
% The distribution of $(Y, U)$ given $T=t$  and $X=x$ is multivariate Gaussian.
% \end{assumption}
%
% \noindent Assumption \ref{asm:copula} implies the following model:
\begin{align}
    & U = \epsilon_U \label{eqn:u}\\
    &T = f_\epsilon(X, U) 
 \label{eqn:treatment_general,multi-y}\\
    % &E[U\mid T=t, X=x] =\mu_{u\mid t,x},\quad Cov(U\mid T=t, X=x) = \Sigma_{u \mid t,x} \label{eqn:conditional-confounder,multi-y} \\
    &Y = g(T,X) + \Gamma_{t, x}\Sigma_{u|t,x}^{-1/2}U + \epsilon_Y, \label{eqn:epsilon_y}
\end{align}
where $U$ is mean zero and has identity covariance without loss of generality and we define $\Sigma_{u \mid t, x} := \text{Cov}(U \mid T=t, X=x)$. $\epsilon_Y$ is mean zero with diagonal covariance $\Delta$ independent of $T, X$ and $U$. 
The proposed structural model satisfies \eqref{eqn:obs_moments} and implies that $E[Y | do(T=t), X=x] := g(t, x) = \check g(t, x) - \Gamma_{t,x}\Sigma_{u|t,x}^{-1/2}\mu_{u\mid t, x}$ is the intervention mean, where $\mu_{u\mid t, x} := E[U|T=t,X=x]$. The intervention mean differs from the observed data mean by an unidentifiable bias, $\Gamma_{t,x}\Sigma_{u|t,x}^{-1/2}\mu_{u\mid t, x}$.  We say that $U$ are \emph{potential} confounders because they are upstream of the treatment and outcomes, but are only truly confounding if $\Gamma_{t,x}\Sigma_{u|t,x}^{-1/2}\mu_{u\mid t, x}$ is non-zero.

% The structural equations \eqref{eqn:u}-\eqref{eqn:epsilon_y} imply that the latent variables $U$ are

\begin{definition}[Potential confounding]
\label{asm:potential_conf}
$U$ are \emph{potential confounders} in that they are possible causes of $T$ and $Y$. Further, $T$ and $Y$ are not causes for any function of $U$.
\end{definition}

We cannot test whether latent variables $U$ from the outcome model are potential confounders without the structural equation assumption, since the observed moments in \eqref{eqn:obs_moments} are also consistent with a model in which the latent variables are mediators caused by the treatment\footnote{See \citet{zhang2022interpretable}, who derive omitted variable bounds for the direct and indirect effects in a mediation analysis under a similar framework to the one used in this paper.}.  However, even if $U$ are not potential confounders, our sensitivity analysis will yield conservative bounds on the treatment effects (see Appendix \ref{sec:relaxing}).  

%We include $\Sigma_{u\mid t,x}^{-1/2}$ in (\ref{eqn:epsilon_y}) without loss of generality because $\Gamma_{t,x}$ are the factor loading matrices in the observed outcome regression and simplifies the following results. 

%In contrast, when there is confounding, the mean of the intervention distribution is

When model \eqref{eqn:u}-\eqref{eqn:epsilon_y} holds, the sensitivity analysis is well-defined in that the PATEs are identified given the parameters $\psi_T=\{\mu_{u|t,x}, \Sigma_{u|t,x}: t \in \mathcal{T}, x \in \mathcal{X}\}$, which govern the relationship between confounders and the treatment, and $\psi_Y = \{\Gamma_{t,x} : t \in \mathcal{T}, x \in \mathcal{X}\}$, which are the factor loading matrices governing the influence of the unmeasured confounders on each outcome. 

\subsection{Establishing Sensitivity Bounds}
\label{sec:bias_nonGaussian}

To complete our model specification, we establish an interpretable sensitivity parameterization for $\psi_T=\{\mu_{u|t,x}, \Sigma_{u|t,x}: t \in \mathcal{T}, x \in \mathcal{X}\}$. Following  \citet{cinelli2020making}, we propose a parametrization which imposes no restrictions on the observed data and reflects the strength of linear dependence between $T$ and $U$ via the partial correlation, generalized to account for multivariate $U$.  We assume $U$ is linear in $T$, has constant variance, and is uncorrelated with observed covariates. Let $\rho$ be the $m$-dimensional sensitivity vector corresponding to the partial correlation vector between $T$ and $U$ after regressing out $X$ so that
\begin{align}
\mu_{u\mid t,x} &=  \rho\left(t-\mu_{t\mid x}\right)/\sigmat \label{eqn:conditional_u_mean}, \\
\SigmaU := \Sigma_{u \mid t,x} &= I_m-\rho \rho^{\prime}
\label{eqn:conditional_u_cov}
\end{align}
for all $t \in \mathcal{T}, x \in \mathcal{X}$, where we define $\mu_{t\mid x} := E[T\mid X=x]$ and $\sigmat^2 := Var(T^{\perp X}) = E_X[Var(T\mid X)]$. Equation \eqref{eqn:conditional_u_cov} is implied by Equation \eqref{eqn:conditional_u_mean} and the constraint that $Cov(U) = I$. Since $\Sigma$ is positive definite, we have that $||\rho||^2_2 < 1$. To maintain consistency with related single outcome sensitivity analyses \citep{cinelli2020making}, we denote $\rsqtx = ||\text{Cor}(T^{\perp X}, U^{\perp X})||_2^2 = ||\rho||^2_2$ to be the squared norm of the partial correlation vector between $T$ and $U$ given $x$, which is identically the partial R-squared based on a linear fit to $T$.

% Equations \eqref{eqn:conditional_u_mean}-\eqref{eqn:conditional_u_cov} are implied when $U$ and $T$ are jointly multivariate Gaussian given $x$, but we do not require such explicit assumptions about the distribution of $T$. 

%This parameterization also implies that $E[U\mid X=x] = 0$, $Cov(U)=I_m$, and $Cor(X, U)=0$. 
% Note that $I_m-\frac{\beta \beta^{\prime}}{\sigmat^{2}}$ is only positive definite, if $||\beta||_2^2 < \sigmat^2$, and with a slight abuse of notation, we still use the notation $0
% \leq \rsqtx := \frac{||\beta||^2_2}{\sigmat^2} < 1$ to reflect a measure of the strength of dependence between confounders and the treatment.  

Finally, for the remainder of the paper, we focus on outcome models for which there are no interactions between unobserved confounders and the covariates or the treatment. In the no-interaction model, the residual outcome covariance is constant in $t$ and $x$ and thus is equivalent to the following assumption.

\begin{assumption}[Homoscedasticity]
\label{asm:homoscedasticity}
$\text{Cov}(Y | T=t, X=x)$ is invariant to $t$ and $x$. 
\end{assumption}

\noindent Together, Assumption \ref{asm:homoscedasticity} and sensitivity parameterization  \eqref{eqn:conditional_u_mean}-\eqref{eqn:conditional_u_cov} imply that the factor loadings $\Gamma_{t,x} = \Gamma$ are also invariant to $t$ and $x$.\footnote{We consider extensions for the heteroscedastic outcome model in which $\Gamma$ can vary with $T$ and $X$ in Appendix \ref{sec:relaxing}. } With these additional assumptions, we have the following bound on the treatment effect for all outcomes:

\begin{theorem}
\label{thm:ignorance_region_general,multi-y}
    Assume model \eqref{eqn:u}-\eqref{eqn:epsilon_y} with $\psi_T$ defined by \eqref{eqn:conditional_u_mean}-\eqref{eqn:conditional_u_cov}, and Assumptions \ref{asm:latent-unconfoundedness-scm}-\ref{asm:homoscedasticity}, and let $\sigma^2_{a'Y} := Var(a'Y \mid T=t, X=x)$. The partial fraction of outcome variance explained by confounders is $\rsqytx := ||a'\Gamma||/\sigma^2_{a'Y}$.  The confounding bias of $\text{PATE}_{a,t_1,t_2}$, $\text{Bias}_{a,t_1,t_2}$ is equal to $\frac{a' \Gamma \SigmaU^{-1/2}\rho}{\sigmat}(t_1 - t_2)$ and it is bounded by
    \begin{equation}
    \label{eqn:worst-case-bias-general,multi-y}
        \text{Bias}_{a,t_1,t_2}^2 
        \, \leq \, 
        \frac{(t_1-t_2)^2}{\sigmat^2} \left(\frac{\rsqtx}{1 - \rsqtx}
        \right)\parallel a' \Gamma\parallel_2^2 
        \, \leq \,
        \frac{(t_1-t_2)^2}{\sigmat^2} \left(\frac{\rsqtx}{1 - \rsqtx}
        \right) \sigmat^2_{a'Y}.
    \end{equation}
The first bound is achieved when $\rho$ is collinear with $a'\Gamma$ and the second bound is achieved when $\rsqytx = 1$.
\end{theorem}

\noindent This theorem implies that the true treatment effect for contrast $a'Y$ lies in the interval $a'E_X[\check g(t_1, x) - \check g(t_2, x)] \pm \frac{t_1-t_2}{\sigmat}\sqrt{\frac{\rsqtx}{1 - \rsqtx}}||a' \Gamma||_2$ and as a consequence, for any $\rsqtx$, the bias for outcome $j$ is proportional to the norm of the $j$th row of $\Gamma$.

In the following corollary, we establish a global bound on the biases over all the outcome contrasts $a'Y$ with $||a||_2=1$.

\begin{corollary}
\label{cor:ignorance_region_global,multi-y}
    Let $d_1$ be the largest singular value of $\Gamma$. For all unit vectors $a$, the confounding bias, $\text{Bias}_{a,t_1,t_2}$, is bounded by
    \begin{equation}
    \label{eqn:worst-case-bias-gaussian-global,multi-y}
        \text{Bias}_{a,t_1,t_2}^2 \leq \frac{(t_1-t_2)^2}{\sigmat^2} \frac{\rsqtx}{1 - \rsqtx}d_1^2,
    \end{equation}
    with equality when $a = u_1^{\Gamma}$, the first left singular vector of $\Gamma$, and when $\rho$ is collinear with $v_1^{\Gamma}$, the first right singular vector of $\Gamma$. There is no confounding bias for the causal effect estimates of outcome $a'Y$ when $a \in Null(\Gamma')$.
    % \noindent Proof. See appendix.
\end{corollary}

\noindent For $a = u_1^{\Gamma}$, $a'Y$ corresponds to the linear combination of outcomes that is most correlated with confounders.  In contrast, when $a$ is in the null space of $\Gamma'$, $\text{PATE}_a$ is identified because $a'Y$ is uncorrelated with $U$.

\subsection{Factor Confounding}

%% Having estbalished bounds, we now consider additional assumptions which sharpen/simplify
Under certain additional assumptions about the factor model, we can identify $\Gamma$ up to rotation, so that the bounds in Theorem \ref{thm:ignorance_region_general,multi-y} and Corollary \ref{cor:ignorance_region_global,multi-y} only depend on the single unidentified sensitivity parameter, $\rsqtx$.

\begin{fact}
\label{thm:gamma-identifiability}
Assume outcome model \eqref{eqn:epsilon_y} with homoscedastic residuals (Assumption \ref{asm:homoscedasticity}).  If $\Gamma$ is rank $m$ and there remain two disjoint matrices of rank $m$ after deleting any row of $\Gamma$, then $\Gamma$ is identifiable up to rotations from the right \citep{anderson1956statistical}.
% 
% \noindent Proof. See appendix.
%and we can assume $Cov(U | X=x)=I_m$ without loss of generality
\end{fact}

The bound in Theorem \ref{thm:ignorance_region_general,multi-y} depends on $\Gamma$ only through $||a'\Gamma||_2$ which is invariant to rotations. Likewise, the bound in Corollary \ref{cor:ignorance_region_global,multi-y} depends on the first singular value of $\Gamma$ which is rotation-invariant.  In order for $\Gamma$ to be identified up to rotation, we must have  $(q-m)^2-q-m\geq0$ and each confounder must influence at least three outcomes \citep{anderson1956statistical}. We group the conditions under which the factor model \eqref{eqn:u}-\eqref{eqn:epsilon_y} yields a sensitivity analysis entirely parameterized by $\psi_T = \rho$ into the following assumption.

\begin{assumption}[Factor confounding]
\label{asm:fact_conf}
The proposed causal model satisfies factor confounding. We say that a causal model satisfies factor confounding if the outcomes follow the model proposed in equation \eqref{eqn:epsilon_y}, $U$ are potential confounders (Definition \ref{asm:potential_conf}), and $\Gamma$ is identifiable up to rotations. We say that a model satisfies factor confounding \emph{for outcome $a'Y$} if $||a'\Gamma||_2$ is identifiable. Factor confounding for outcome $a'Y$ implies that the partial fraction of outcome variance explained by confounders, $\rsqytx$, is identifiable.
\end{assumption}

% \begin{assumption}[Factor confounding]
% \label{asm:fact_conf}

% \end{assumption}

 There are some useful ways that practitioners can reason about the plausibility of factor confounding.
In particular, since factor confounding is violated if there are confounders that influence fewer than three outcomes, practitioners should consider carefully whether there are important unmeasured confounders that might influence only one or two outcomes.  While the bulk of our analysis is done under the factor confounding assumption, even when factor confounding is violated, we can still apply our sensitivity analysis, albeit with more conservative bounds on the causal effects.  As such, we view factor confounding as a useful ``reference assumption'' that can help establish informative bounds on the causal effects. For now, we assume factor confounding, and explore additional relaxations of Assumption \ref{asm:fact_conf} in Appendix \ref{sec:relaxing}.  

% In this case, the exact bias depends on the unidentifiable rotations of the factor loading matrices, and as such, we provide a looser bound on the omitted variable bias.

% is required, e.g. one in which $\beta$ can vary with $X$ (See Appendix ??).  
% In contrast to the analogous results for multi-treatment inference \citep{zheng2021copula}, the bias for treatment $a'Y$ is unbounded, since $\frac{\rsqtx}{1 - \rsqtx}$ can be arbitrarily large.  

\subsection{Null Control Outcomes}
\label{sec:nco}

In this section, we establish how additional assumptions about null control outcomes constrain the shared sensitivity vector, $\rho$, and thus reduce the size of the partial identification regions for the causal effects of the other outcomes.  Under null control assumptions, we fix the biases for a set of so-called null control outcomes to match the observed effect under NUC, so that the causal effects for the null control outcomes are zero after accounting for confounding biases.  Such assumptions add important context, in particular because Theorem \ref{thm:ignorance_region_general,multi-y} only establishes marginal bounds for any treatment contrast, but does not account for the dependence in the omitted variable bias across outcomes.  Null control assumptions reveal the joint relationship between biases across outcomes.  Our results complement those from \citet{rosenbaum2021sensitivity}, who explore sensitivity analysis for matching estimators and demonstrate that null control outcomes can make tests of significance on the primary outcome more robust to confounding. 

For a fixed treatment contrast $t_1$ versus $t_2$, and with a slight abuse of notation, we let $\tau$ correspond to the $q$-vector of PATEs on each of the measured outcomes and let $\teNaive$ denote the $q$-vector of PATEs under NUC. Let $\mathcal{C} \subset \{1, \dots, q\}$ be a set of indices for $c < q$ null control outcomes for which there is assumed to be no causal effect of the treatment on these $c$ measured outcomes, that is, $\tau_j = 0$ for any $j \in \mathcal{C}$. For these null control outcomes, the corresponding c-vector of treatment effects under NUC, $\teNaiveNull$, must equal the corresponding confounding biases. Since the bias is a function of the sensitivity vector $\rho$, we have that $\teNaiveNull = \frac{t_1-t_2}{\sigma\sqrt{1 - \rsqtx}} \Gamma_{\mathcal{C}} \rho$
where $\Gamma_{\mathcal{C}}$ is a $c \times m$ matrix equal to the $c$ rows of $\Gamma$ corresponding to null control outcomes. This equation implies that $\teNaiveNull$ must be in the column space of $\Gamma_{\mathcal{C}}$ and also implies a lower bound on the fraction of confounding variation in the treatment, $\rsqtx$. 

\begin{proposition}
\label{prop:cali_wnco}
Assume model \eqref{eqn:u}-\eqref{eqn:epsilon_y}  with sensitivity parameterization \eqref{eqn:conditional_u_mean}-\eqref{eqn:conditional_u_cov} and Assumptions \ref{asm:latent-unconfoundedness-scm}-\ref{asm:homoscedasticity}. Further, suppose there are $c$ null control outcomes, $Y_j$, such that $\tau_j = 0$ for $j \in \mathcal{C}$. Then, $\teNaiveNull$ must be in the column space of $\Gamma_{\mathcal{C}}$. In addition, the fraction of variation in the treatment due to the confounding is lower bounded by
\begin{equation}
    \rsqtx \geq R_{\text{min}, \mathcal{C}}^2 := \frac{\sigmat^2 \parallel 
    \Gamma_{\mathcal{C}}^{\dagger} \teNaiveNull \parallel_2^2}{(t_1-t_2)^2 + \sigmat^2 \parallel \Gamma_{\mathcal{C}}^{\dagger} \teNaiveNull \parallel_2^2},
\end{equation}
where $\Gamma_{\mathcal{C}}^{\dagger}$ denotes the pseudoinverse of $\Gamma_{\mathcal{C}}$. $R_{\text{min}, \mathcal{C}}^2$ is identifiable under factor confounding (Assumption \ref{asm:fact_conf}).
% Proof: See Appendix.
\end{proposition}

\noindent When the number of null control outcomes is smaller than the rank of $\Gamma$, then $\teNaiveNull$ is automatically in the column space of $\Gamma$.  In order to correct for the biases of null control outcomes, confounding must explain at least $R^2_{min, \mathcal{C}}$ of the residual treatment variance.  Moreover, for any assumed $\rsqtx \geq R^2_{min, \mathcal{C}}$, the assumption of null controls constrains the space of possible effects for the non-null outcomes.  We formalize this below. 
% Gamma_C can be rotated to be full rank.

%exactly $R^2_{min}/\rsqtx$ of the residual treatment variance that is due to confounding can be attributed to confounders that are shared with the null controls.  This constrains the magnitude of the remaining confounding bias. Below, we formalize precisely how the null controls assumptions influence the bias on other outcomes.

\begin{theorem}
\label{thm:ignorance-region-gaussian-wnc,multi-y}
Under the assumptions established in Proposition \ref{prop:cali_wnco}, for any value of $\rsqtx \geq R^2_{min, \mathcal{C}}$, the confounding bias for the treatment effect on outcome $a'Y$ is in the interval
\begin{equation}
    \label{eqn:ignorance-region-gaussian-wnc,multi-y}
    \text{Bias}_{a,t_1,t_2} \in 
    \left[a' \Gamma \Gamma_{\mathcal{C}}^{\dagger} \teNaiveNull
    \; \pm \;
    \parallel
    a'\Gamma P_{\Gamma_{\mathcal{C}}}^{\perp} 
    \parallel_2
    \sqrt{
    \frac{(t_1-t_2)^2}{\sigmat^2}\left(
    \frac{\rsqtx}{1 - \rsqtx} - 
    \frac{R^2_{min, \mathcal{C}}}{1 - R^2_{min, \mathcal{C}}} 
    \right)}\,
\right],
\end{equation}
where $P_{\Gamma_{\mathcal{C}}}^{\perp} = I_m - \Gamma_{\mathcal{C}}^{\dagger} \Gamma_{\mathcal{C}}$ is the $m \times m$ projection matrix onto the space orthogonal to the row space of $\Gamma_{\mathcal{C}}$. Under Assumption \ref{asm:fact_conf}, $\rsqtx$ is the only unidentifiable parameter.
% Proof: See Appendix.
\end{theorem}
\noindent Note that the ignorance region is no longer centered at $a'\teNaive$ but instead $a'\teNaive - a' \Gamma \Gamma_{\mathcal{C}}^{\dagger} \teNaiveNull$, where $a' \Gamma \Gamma_{\mathcal{C}}^{\dagger} \teNaiveNull$ is the bias correction under the null controls assumption. Theorem \ref{thm:ignorance-region-gaussian-wnc,multi-y} indicates that whenever $\Gamma_{\mathcal{C}}$ is of rank $m$ or whenever we assume $\rsqtx = R^2_{min, \mathcal{C}}$, treatment effects for all outcomes are identifiable under factor confounding. A direct comparison of the ignorance regions from Theorem \ref{thm:ignorance_region_general,multi-y} and Theorem \ref{thm:ignorance-region-gaussian-wnc,multi-y} indicates that after incorporating null control outcomes, for any fixed $\rsqtx$ the width of the ignorance region is reduced by a multiplicative factor of 
%
% \begin{corollary}
% \label{cor:ignorance_width}
% Under assumptions established in Theorem \ref{thm:ignorance-region-gaussian-wnc,multi-y}, null control outcomes reduce the width of the ignorance region by a multiplicative factor of 
\begin{equation}
    \sqrt{
    1 - \left(\frac{R^2_{min, \mathcal{C}}}{1 - R^2_{min, \mathcal{C}}} \bigg/ \frac{\rsqtx}{1 - \rsqtx}
    \right)}
    \frac{\parallel
    a'\Gamma P_{\Gamma_{\mathcal{C}}}^{\perp} 
    % (I - \Gamma_{\mathcal{C}}^{\dagger}\Gamma_{\mathcal{C}}) \Gamma' a
    \parallel_2}{
    \parallel a'\Gamma \parallel_2}
    \leq 1.
\label{eqn:ignorance_width}
\end{equation}
%\end{corollary}
From Equation \eqref{eqn:ignorance_width}, it is evident that null controls reduce the width of the worst-case ignorance region in two ways. The first factor under the radical is due to the fact that only $\rsqtx - R^2_{min, \mathcal{C}}$ of the treatment variance can be due to confounders which are uncorrelated with the null control outcomes.  This factor reduces the width of the ignorance regions for all non-null outcomes by an equal proportion. As a special case, when all the unobserved confounders are correlated with the null control outcomes, then $\rsqtx = R^2_{min, \mathcal{C}}$ and the treatment effects for all outcomes are identified.  In contrast, the second factor depends on the specific outcome of interest, $a'Y$. The ignorance region shrinks the most for outcomes that are mostly correlated with the same set of confounders as the null control outcomes. Mathematically, when $a'\Gamma$ is in the row space of $\Gamma_{\mathcal{C}}$, the treatment effect of $a'Y$ is identified under factor confounding. When $a'\Gamma$ is orthogonal to the row space of $\Gamma_{\mathcal{C}}$, $\frac{\parallel a'\Gamma P_{\Gamma_{\mathcal{C}}}^{\perp} \parallel_2}{\parallel a'\Gamma \parallel_2} = 1$, so that the confounders affecting the null control outcomes are independent of the confounders affecting $a'Y$, and thus there is no further reduction of the ignorance region. In summary, the best null control outcomes are those which have large confounding biases (and hence large values of $R^2_{min, \mathcal{C}}$) and also have similar outcome-confounder associations with the other outcomes of interest. We illustrate these facts in a simulation study in Section \ref{sec:simulation}.

\subsection{Robustness}

\label{sec:cal_rv}

A common strategy for characterizing the robustness to confounding is to identify the ``smallest'' sensitivity parameter(s) which nullifies the causal effect.  For example, for single-outcome inference \citet{cinelli2020making} define the robustness value, $RV^1_a$, as the smallest value of $\rsqtx$, assuming $\rsqtx = \rsqytx$, needed to change the sign of the effect\footnote{Here we use the superscript  1, to emphasize that this is a robustness value for single outcome analyses.} and define an ``extreme robustness value'' ($XRV_a$) as the smallest value of $\rsqtx$, assuming  $R^2_{a'Y\sim U \mid T, X}=1$, needed to change the sign \citep{cinelli2022iv}. The $XRV_a$ is a more conservative measure of robustness than $RV^1_a$ since it assumes that all the residual outcome variance is attributable to confounders. 

Here, we define the factor confounding robustness value, $RV^\Gamma_a$, for outcome $a'Y$ as the smallest value of $\rsqtx$ needed to make the causal effect zero under factor confounding. $RV^\Gamma_a$ is a more accurate reflection of robustness than $RV^{1}_a$ or $XRV_a$ when factor confounding holds, because $\rsqytx$ is identified under this assumption for all $a$ (Assumption \ref{asm:fact_conf}). $XRV_a \leq RV^\Gamma_a < 1$ and $RV^\Gamma_a$ can be either smaller or larger than $RV^1_a$, depending on how much variance is attributed to potential confounders from the factor model. When $\rsqytx = ||a'\Gamma||_2^2/(||a'\Gamma||_2^2 + a'\Delta a) \geq RV^1_a$, then $RV^\Gamma_a \leq RV^1_a$.  Conversely, when $\rsqytx < RV^1_a$, then $RV^\Gamma_a > RV^1_a$. $RV^\Gamma_a = XRV_a$ if and only if $a'\Delta a = 0$, that is, the latent factors explain all the residual outcome variance for $a'Y$.  We demonstrate the relationship between these robustness values in a simulation study in Section \ref{sec:simulation}. 

% Under the assumptions in Theorem \ref{thm:ignorance_region_general,multi-y}, $RV^{\Gamma}_a$ is available analytically as $RV^{\Gamma}_{a} = \frac{\omega}{1+\omega}$ with $\omega = \left(\frac{a'\teNaive \sigma}{(t_1-t_2) \parallel a'\Gamma \parallel_2}\right)^2$ and where $\teNaive$ denotes the vector of PATEs for all outcomes under NUC (see Corollary \ref{corollary:rv_increase}).

In addition, we can quantify how assumptions about null control outcomes influence $RV_a^\Gamma$. First, note that when there is only a single null control, indexed by $c$, we have that $R^2_{min, \mathcal{C}}= RV^{\Gamma}_{e_c}$, since $RV^{\Gamma}_{e_c}$ is the smallest fraction of treatment variance needed to nullify outcome $c$. In other words, the total fraction of treatment variance explained by confounders is lower bounded by the robustness value for the null control.  We define the ``combined robustness value'' with null controls, $\mathcal{C}$, as 
\begin{equation}
\label{def:combined-rv}
RV_{a,\mathcal{C}}^\Gamma :=  \underset{\rho}{min}\,\, ||\rho||^2_2 \text{ s.t. } a'\tau = 0 \text{ and } e_c'\tau = 0, \,\forall\, c \in \mathcal{C},
\end{equation}

\noindent where $e_c$ is the $c$th canonical basis vector.  $RV_{a,\mathcal{C}}^\Gamma$ corresponds to the minimum fraction of treatment variance explained by unobserved confounders that is required to make the causal effect on $a'Y$ equal to zero \emph{and} satisfy all null control assumptions. Naturally, the minimum fraction of treatment variance explained by confounders needed to satisfy the null control assumptions and nullify $a'Y$ must be larger than the minimum fraction needed to nullify just $a'Y$.  

\begin{theorem}
\label{corollary:rv_increase}
Let $\teNaive$ denote the vector of PATEs for all outcomes and $\teNaiveNull$ be the vector of PATEs for null control outcomes under NUC. Under the assumptions established in Proposition \ref{prop:cali_wnco}, the factor confounding robustness value for outcome $a'Y$ is
\begin{equation}
RV^{\Gamma}_{a} = \frac{\omega}{1+\omega},
\end{equation}
\noindent where $\omega = \frac{\sigmat^2}{(t_1-t_2)^2}\frac{(a'\teNaive)^2}{ \parallel a'\Gamma \parallel_2^2}$ . The combined robustness value for outcome $a'Y$ given null controls $\mathcal{C}$ is
\begin{equation}
\label{eqn:rv_combined}
RV_{a,\mathcal{C}}^\Gamma = \frac{w_{\mathcal{C}}}{1 + w_\mathcal{C}}\geq max(R^2_{min, \mathcal{C}}, RV^{\Gamma}_{a}),
\end{equation}
where $w_{\mathcal{C}} := \frac{\sigma^2}{(t_1-t_2)^2} \left[\frac{(a'(\teNaive - \Gamma\Gamma^{\dagger}_{\mathcal{C}}\teNaiveNull))^2}{\parallel a'\Gamma P_{\Gamma_{\mathcal{C}}}^{\perp} \parallel^2_2} + \parallel \Gamma_{\mathcal{C}}^{\dagger}\teNaiveNull\parallel^2_2\right]$.
% Proof: See Appendix.
\end{theorem}
\noindent In the following examples, in addition to the combined robustness value $RV^\Gamma_{a, \mathcal{C}}$, we report another useful summary, $RV^\Gamma_{a, \mathcal{C}} - R^2_{min, \mathcal{C}}$ which corresponds to the additional fraction of variance explained by confounders that is needed to nullify outcome $a'Y$ beyond the amount needed to nullify the null control outcomes, $\mathcal{C}$.

\section{Simulation Study}
\label{sec:simulation}
In this section, we provide intuition for the theoretical results in a simple simulated example without covariates where both the treatment and outcome model are linear and Gaussian.  We generate $n=1000$ observations from model \eqref{eqn:u}-\eqref{eqn:epsilon_y} with $m=2$ latent variables, $q=10$ outcomes and fix $T=f_\epsilon(U) = \beta' U + \epsilon_T$ in \eqref{eqn:treatment_general,multi-y} and $g(t) = \tau t$ in \eqref{eqn:epsilon_y}, where $\te$ represents the $q-$vector of causal effects for a unit change in $t$.  We choose $\te$ so that there is no causal effect on the first, second, and tenth outcomes, and a causal effect of one for the other seven outcomes.  

\begin{figure}[t!]	
	\centering
	\begin{subfigure}[t]{0.2\textwidth}
		\centering
		\includegraphics[width=\textwidth]{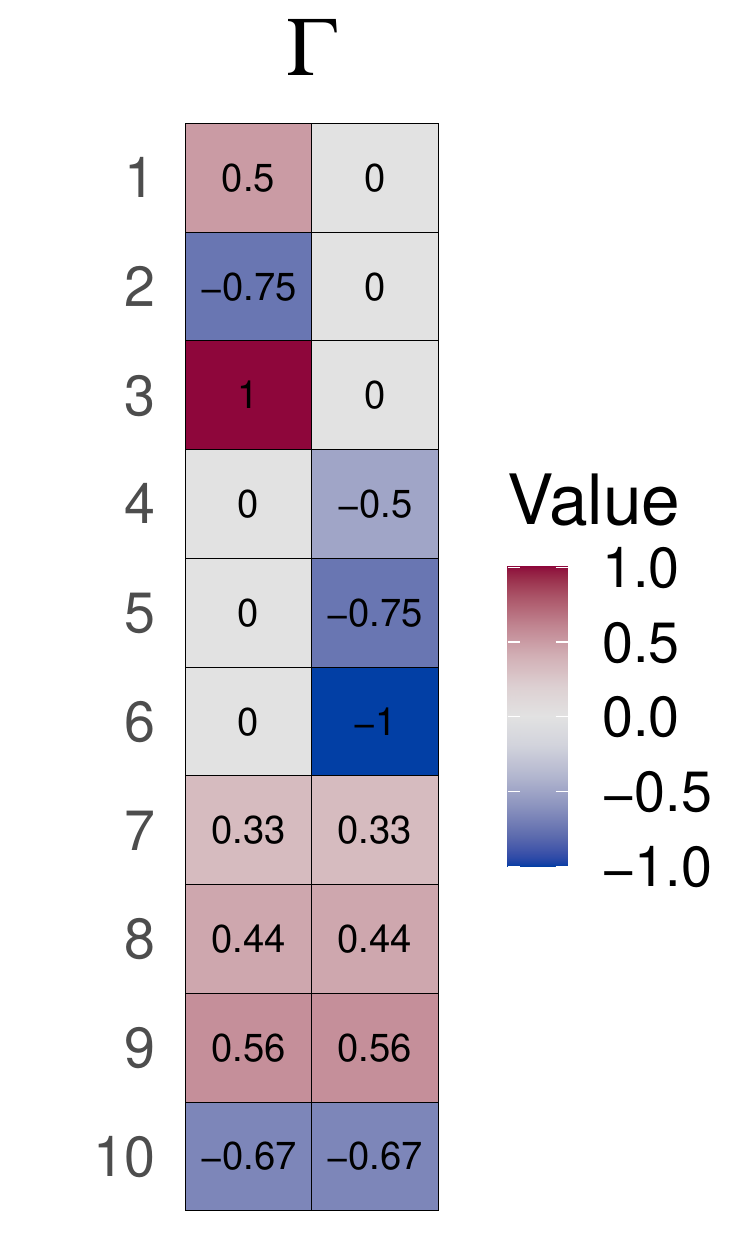}
		\vspace*{0pt}
		\caption{\label{fig:gamma_heat}}	
	\end{subfigure}
	\hspace{.25in}
    \begin{subfigure}[t]{0.7\textwidth}
	    \centering
	    \includegraphics[width=\textwidth]{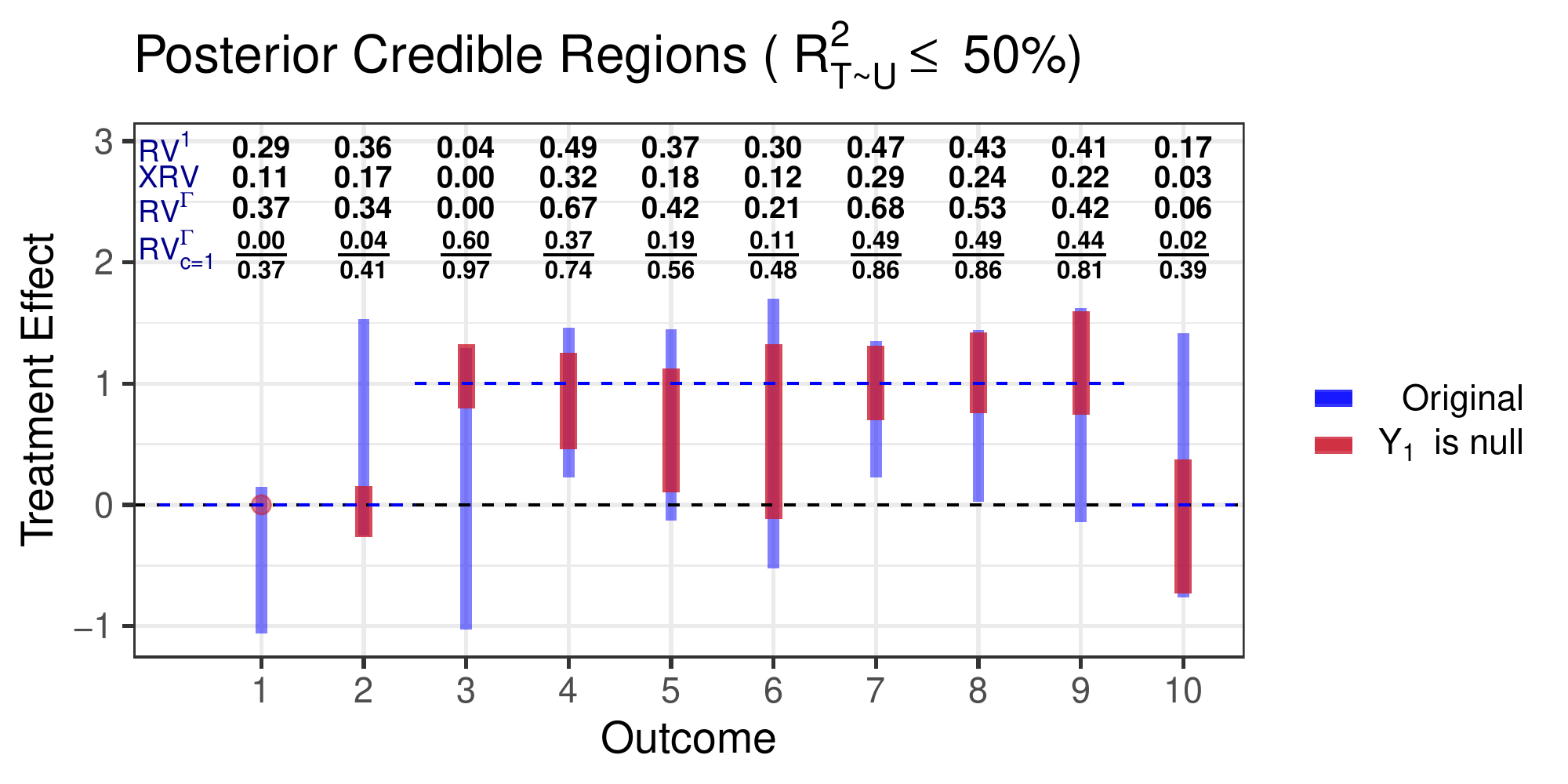}
	    \caption{\label{fig:sim_intervals}}
    \end{subfigure}
    	 \caption{a) Heatmap of the $\Gamma$. b) 95\% posterior credible regions for the causal effects on each of the ten outcomes for $R^2_{T\sim U} \leq 0.5$ based on $n=1000$ simulated observations without any null controls (blue) and with $Y_1$ as a null control (red).    With the null control assumption, outcomes 2 and 3 are identifiable because the corresponding rows of $\Gamma$ are collinear with row 1.  Rows 4-6 of $\Gamma$ are orthogonal to the first row of $\Gamma$, so there is only a small reduction in the size of the identification region and no change in the midpoint of this region.  Multiple robustness values are reported above the intervals, including the single outcome robustness ($RV^1$), extreme robustness ($XRV$), and robustness under factor confounding ($RV^\Gamma$).  The row labeled $RV^\Gamma_c$ summarizes the robustness under factor confounding with null controls, where the denominator is the combined robustness value defined in \eqref{eqn:rv_combined} and the numerator is the fraction of outcome variance needed to nullify outcome $Y_i$ beyond that induced solely by the null control, $RV^{\Gamma}_{i,c} - R^2_{min, c}$. The null control assumption significantly increases the robustness of all non-null effects, and is enough to effectively nullify outcomes $2$ and $10$.
    	 \label{fig:sim_plots}}
\end{figure}

The partial correlation vector between $T$ and $U$, $\rho = \frac{\beta}{\text{Var}(T)}$, is chosen uniformly on a sphere with $||\rho||_2^2 = R^2_{T\sim U} = 0.5$.  We also choose $\Gamma$ with a particularly simple structure, shown in Figure \ref{fig:gamma_heat}.  After generating data, we infer $\Gamma$ and $\teNaive$ using a Bayesian multivariate linear regression model with a factor model structure on the residual covariance using the probabilistic programming language Stan \citep{stan}. 

% The Bayesian model accounts for uncertainty in $\teNaive$ as well as uncertainty in $\Gamma$.  

In Figure \ref{fig:sim_intervals}, we plot the 95\% posterior credible interval for the effect of one unit change in $t$ on each outcome assuming $R^2_{T\sim U} \leq 0.5$ (blue) by computing the relevant posterior quantiles of the endpoints of the partial identification region determined by Theorem \ref{thm:ignorance_region_general,multi-y}.  The outcomes with the largest intervals are those for which the corresponding rows of $\Gamma$ have the largest magnitudes (i.e. darker colors in Figure \ref{fig:gamma_heat}).  Only outcomes 4, 7, and 8 are robustly different from zero at $\rsqt \leq 0.5$.

% These robustness values are contingent on ..., but when this assumption holds, are more informative than the single outcome robustness values proposed by \citet{cinelli2020making}.  In Appendix Figure \ref{}, we compare the multi-outcome robustness values (Equation \ref{}) to the single outcome robustness value, for which .  Some outcomes have larger robustness values than the single outcome RV whereas some have smaller robustness values, depending on $||\Gamma||_2$.  In fact, the multi-outcome RV can be arbitrarily close to zero (as the norm of the corresponding row of $\Gamma$ shrinks toward 0) or arbitrarily close to 1 (as the norm of the corresponding row of $\Gamma$ grows), independent of the single outcome robustness value.

We then trace the implications of a null control assumption through the sensitivity model to illustrate the results of Section \ref{sec:nco}. In Figure \ref{fig:sim_intervals} we plot the posterior 95\% credible regions under the additional assumption that the first outcome is a null control outcome (red).  After incorporating the null control assumption, the posterior credible regions for outcomes 2-10 still include the true causal effects assuming factor confounding and $\rsqt \leq 0.5$. Among the non-null outcomes, $Y_6$ is the only outcome with an ignorance region that still includes zero. Since the first three rows of $\Gamma$ are mutually collinear, fixing the bias of $Y_1$ implies that, with infinite data, the effects for $Y_2$ and $Y_3$ are also identified since $\Gamma_{i}P_{\Gamma_\mathcal{C}}^\perp = 0$ for rows $i \in \{2, 3\}$ (Equation \eqref{eqn:ignorance_width}). Rows 4-10 are not mutually collinear with the first row of $\Gamma$ and thus, even with infinite data, the effects remain unidentified. Rows 4-6 of $\Gamma$ are orthogonal to row 1, which implies that $\Gamma_{i}P_{\Gamma_\mathcal{C}}^\perp = \Gamma_{i}$ for $i \in \{4, 5, 6\}$ and thus the midpoints of the intervals remain unchanged for outcomes 4-6.  The interval widths still shrink slightly, since only $\rsqt-R^2_{min,c} = 0.5 - 0.37 = 0.13$ of the treatment variance can be explained by the confounders of $Y_4$ through $Y_6$ after accounting for the null control outcomes. The dot products of $\Gamma_1$ with rows $7$ through $10$ are all nonzero, which means that the midpoints of the ignorance regions change for outcomes $7$ through $10$ after incorporating the null control assumption. Consistent with \eqref{eqn:ignorance-region-gaussian-wnc,multi-y}, the directional change in the midpoint of ignorance region for outcome $Y_i$ is determined by the sign of the dot product of row $\Gamma_1$ and $\Gamma_i$ (positive for $Y_7$ to $Y_9$ and negative for $Y_{10}$).

% The sign of dot product determines the sign of the bias correction .

Finally, we compare four different measures of robustness to confounding in black above the corresponding intervals: single outcome robustness ($RV^1$) \citep{cinelli2020making}, extreme robustness ($XRV$), robustness under factor confounding ($RV^\Gamma$) and robustness under factor confounding with the first outcome as a null control $RV^\Gamma_{c=1}$. By definition, XRV is the most conservative (smallest robustness values). In this simulation, $RV^1$ is smaller than $RV^\Gamma$ for six of the outcomes and larger for the other four. $RV^\Gamma$ is a more accurate reflection of the true robustness under factor confounding since we are able to infer the implied values of $R^2_{Y_i\sim U \mid T}$.  After correctly incorporating the null control assumption, the robustness of all the outcomes with true non-zero effects increases significantly. The largest increase in robustness occurs for $Y_3$, for which the causal effect is identifiable under the null control assumption.  For outcomes $Y_2$ and $Y_{10}$, the other outcomes with no true causal effect, $RV^{\Gamma}_{\mathcal{C}} - R^2_{min,c}$ are close to zero, which means that the null control assumption alone is sufficient to nullify these effects as well.  

% Like $Y_3$, the effect on $Y_2$ is identifiable under the null control assumption since the corresponding rows of $\Gamma$ are collinear.  

% If the posterior interval for an effect includes 0 when $R^2_{T\sim U} = R^2_{min}$ then we simply report `X' in lieu of the robustness value, which means that no additional confounding is needed to explain away the significance of the effect.

\section{Calibration}
\label{sec:calibration,multi-y}

For a sensitivity analysis to be of practical value, it is essential to calibrate the magnitude of the sensitivity parameters against interpretable benchmarks. In this section, we briefly describe strategies for calibrating $\rsqtx$, the sole sensitivity parameter of the worst-case bias under factor confounding. In Section \ref{sec:cali_binary_t}, we propose an alternative sensitivity parameterization tailored to binary treatments which we apply to the analysis in Section \ref{sec:nhanes}.

% and explore approaches for relaxing the factor confounding assumption in Section \ref{sec:cal_r2y}.
% As shown in Theorem \ref{thm:ignorance_region_general,multi-y}, in the absence of additional assumptions, the confounding bias of the NUC estimator for the causal effect on $a'Y$ is maximized at $\text{Bias}_{a,t_1,t_2}^2 = \frac{1}{\sigma_{t\mid x}^2} \left(\frac{\rsqtx}{1 - \rsqtx}\right)\parallel a' \Gamma(t_1-t_2)\parallel_2^2$ when $\beta$ is collinear with $a'\Gamma$.  
% Further, the results from Section \ref{sec:nco} establish the orientation of $\beta$ that maximizes the bounds of the treatment effect given any null control assumptions and any $\rsqtx \geq R^2_{min}$.  As such, we primarily focus on techniques for calibrating and benchmarking the magnitude the scalar parameter $\rsqtx$.

\subsection{Calibrating $\rsqtx$}
\label{sec:cal_rsqtx}

Recall that the magnitude of the sensitivity vector $\rho$ in $\eqref{eqn:conditional_u_mean}$ can be characterized by $\rsqtx$. For linear models, $\rsqtx$ can directly be interpreted as the partial fraction of treatment variance explained by confounders given $X$,
    $\rsqtx = \frac{\text{Var}(T | X) - \text{Var}(T |  U,X)}{\text{Var}(T| X)} = \|\rho \|^2_2$,
and more generally, as the squared norm of the partial correlations between the treatment and confounders (see Section \ref{sec:bias_nonGaussian}). Following prior work, we can calibrate $\rsqtx$ by comparing it to an estimable partial fraction of variance explained. For a reference covariate (or set of covariates), $X_j$, and given all baseline covariates $X_{-j}$, we compute the partial R-squared $R_{T \sim X_j \mid X_{-j}}^2 := \frac{R_{T \sim X}^2 - R_{T \sim X_{-j}}^2}{1 - R_{T \sim X_{-j}}^2}$, which serves as a reference for the magnitude of $\rsqtx$. See \citet{cinelli2020making}, \citet{veitch2020sense} and \citet{zheng2021copula} for additional details on calibration strategies.

%Alternatively, additional implicit information about the direction of $\beta$ can come from assumptions about the magnitude of treatment effects themselves, e.g with the use of null controls or sparsity assumptions \citep[e.g.][]{wang2017confounder}. In Section \ref{sec:cali_wnco}, we discuss the role of null control outcomes, how they constrain both the magnitude and direction of $\beta$, and can strengthen the evidence for or against certain causal effects.

% \subsection{Robustness Values} 

% \noindent There is some subtlety in the definition of $RV_{a,\mathcal{C}}^{\Gamma}$, since $a' \teNaive - a' \Gamma \Gamma_{\mathcal{C}}^{\dagger} \teNaiveNull$ can have opposite sign of $\teNaive$, in which case $RV_{a,{\mathcal{C}}}^\Gamma$ refers to the robustness of the sign of $a' \teNaive - a' \Gamma \Gamma_{\mathcal{C}}^{\dagger} \teNaiveNull$, which is the effect when all the confounding bias is captured by the null controls. 

% \noindent \textbf{Calibration for binary-valued treatments.} 
\subsection{Calibration for Binary Treatments}
\label{sec:cali_binary_t}
For binary-valued treatments, we suggest an alternative sensitivity parameterization which is closely related to the parameterization in the marginal sensitivity model for inference with binary-valued treatments in single outcome problems \citep{tan2006distributional}.  Let $e(X) = P(T=1 \mid X)$ be the ``nominal'' propensity score, $e_0(X, U) = P(T=1 \mid X, U)$ be the ``true'' propensity score, and $\lambda(X, U) = \frac{e_0(X, U)/(1-e_0(X, U))}{e(X)/(1-e(X))}$ be the multiplicative change in the odds of treatment after accounting for unmeasured confounders. The core assumption of the marginal sensitivity model is that the odds of treatment is bounded by some constant, that is, there exists a $\Lambda$ such that $\Lambda^{-1} \leq \frac{e_0(X, U)/(1-e_0(X, U))}{e(X)/(1-e(X))}\leq \Lambda$ holds with probability one. Under parametric assumptions about the conditional distribution of $U$, we can characterize the full distribution of $\lambda(X, U)$. Although not strictly necessary, in this work we focus our examples on settings in which $U$ is assumed to be Gaussian given $T$ and $X$.  

\begin{proposition} 
Assume $U$ is conditionally Gaussian with mean $\muU$ and covariance $\SigmaU$ as given in Equations \eqref{eqn:conditional_u_mean} and \eqref{eqn:conditional_u_cov}.  Further, denote $e(x) = P(T=1\mid X=x)$ and $e_0(x, U) = P(T=1 \mid X=x, U)$.  Then, for any $\rho$ we have $\lambda(X=x, U) = \frac{e_0(x, U)/(1-e_0(x, U))}{e(x)/(1-e(x))} = \text{exp}(V_x)$, where $V_x = (2I_x-1)Z$, $I_x\sim \text{Ber}(e(x)), Z \sim N(\mu_\lambda, \sigma^2_\lambda )$ with $\mu_\lambda = \frac{1}{2\sigma^2}\frac{\rsqtx}{1-\rsqtx}$ and $\sigma^2_\lambda = \frac{1}{\sigma^2}\frac{\rsqtx}{1-\rsqtx}$.
% \noindent Proof. See appendix.
\label{prop:lambda}
\end{proposition}

\noindent Proposition \ref{prop:lambda} states that the log odds ratio, $\text{log}(\lambda(X=x, U))$, is a two-component mixture of normal distributions. Using this proposition, we can find  $\Lambda_\alpha$, such that
%\begin{equation}
$P(\Lambda^{-1}_\alpha \leq \lambda(X, U) \leq \Lambda_\alpha) \geq 1-\alpha.    
\label{eqn:lambda_bound}$
%\end{equation}
Unconditional on $x$, $\lambda(X,U)$ is a two-component mixture with means $\pm \mu_\lambda$, variance $\sigma^2_{\lambda}$ and mixture weights $E[e(X)]$ and $1-E[e(X)]$, which we use to compute $\Lambda_\alpha$.  Since $\mu_\lambda$ and $\sigma^2_{\lambda}$ only depend on $R^2_{T\sim U|X}$ and $\sigma^2$, we can also derive the robustness of outcome $a'Y$ in the $\Lambda$-parameterization by replacing   $R^2_{T\sim U|X}$ with $RV^{\Gamma}_a$ in the formulas for $\mu_\lambda$ and $\sigma^2_{\lambda}$.  Similarly, we can benchmark $\Lambda_\alpha$ by computing how much the odds of treatment changes when adding a reference covariate into a propensity model which already includes some baseline covariates   \citep[see e.g.][]{kallus2021minimax, dorn2022sharp}.  We compute the 1-$\alpha$th quantile of the log odds $\frac{e(X_{base}, X_{ref})/(1-e(X_{base}, X_{ref}))}{e(X_{base})/(1-e(X_{base}))}$ where $X_{base}$ is a set of observed baseline covariates and $X_{ref}$ is a set of reference covariates. We demonstrate an analysis using this benchmarking strategy in the empirical example in the next section.

\section{Analyzing the Effects of Light Alcohol Consumption}
\label{sec:nhanes}
We apply our proposed multi-outcome sensitivity analysis in an investigation of a long-standing question about the potential health benefits of light to moderate alcohol consumption on health outcomes.  In particular, observational data indicates that light alcohol consumption is positively correlated with blood levels of HDL (``good cholesterol'') and negatively correlated with LDL (``bad cholesterol'') \citep[e.g.][]{choudhury1994alcohol, meister2000health, o2007alcohol}.  However, there are known to be many potential confounders related to diet and lifestyle which could explain these associations. We consider treated individuals ($n_T=114$) as those who self-reported drinking between one and three alcoholic beverages per day, and untreated individuals ($n_C=1439$) as those that averaged one drink per week or less.  We make use of laboratory outcomes, $Y = (Y_1, \cdots, Y_{10})$, which consist of three measures of cholesterol (HDL, LDL, and triglycerides), as well as blood levels of potassium, iron, sodium, and glucose and levels of three environmental toxicants, methylmercury, cadmium and lead, all collected from 2017-2020 (pre-pandemic) as part of the National Health and Nutrition Examination Study (NHANES). We control for observed confounders which include age, gender, and an indicator for education beyond a high school degree.  

% We briefly compare our results to a similar analysis on the effects of light drinking of HDL cholesterol by \citet{rosenbaum2021sensitivity}.

% . This approach diverges from \citet{rosenbaum2021sensitivity}, who conduct a sensitivity analysis for the average treatment effect among the treated (ATT) using a significantly reduced dataset of matched pairs. 

In our analysis, we consider the layers of assumptions that one might make to reason about the set of causal effects which are consistent with the observed data.  First, we report posterior intervals for causal effects under the NUC assumption, and then consider bounds on the causal effects under factor confounding (Assumption \ref{asm:fact_conf}).  We calibrate the bounds by benchmarking values of $\Lambda_{0.95}$, the 95th percentile of the multiplicative change in the odds of treatment after accounting for unobserved confounders (Section \ref{sec:cali_binary_t}).  Alternatively, in lieu of specifying $\Lambda_{0.95}$ directly, we also explore the implications of a carefully chosen null control outcome on non-null outcomes.  We report robustness values for all outcomes both with and without the null control assumption. 

% Finally, following the suggestions in Section \ref{sec:cal_r2y}, we use the inferred factor loading matrix to reason about the plausibility of factor confounding for each outcome.  Specifically, we use inferred values of $\rsqytx$ to identify outcomes for which there may be additional confounding that is not reflected in the inferred factor loading matrix.  We focus primarily on potential violations of factor confounding for our chosen null control outcome, for which the inferred value of $\rsqytx$ under factor confounding is relatively small.  

We start by estimating associations under NUC by fitting a multivariate linear regression model with factor-structured residuals, where the estimands reduce to the regression coefficients, $\tau_j$, for $j = 1, \cdots, 10$. The logarithms of all outcomes are approximately unimodal, symmetric, and not heavy-tailed, and thus we regress the log outcomes on age, gender, education, and the alcohol consumption indicator to estimate $\teNaive_j$ under an assumption of no unobserved confounding.  
Here, we assume $g(t) = \tau t$ in \eqref{eqn:epsilon_y} and fit a Bayesian multivariate linear regression with a rank-$m$ factor model structure on the Gaussian residuals using STAN \citep{stan}.  Note that factor confounding can only be satisfied if $m \le 6$ otherwise $(q − m)^2 − q − m < 0$.  As such, we fit models of rank $m \le 6$ and use Pareto-Smoothed Importance Sampling estimates of the leave-one-out cross-validation loss to evaluate relative model fit \citep{vehtari2017practical}. We compare differences in expected log predictive density (ELPD) for different ranks and find that models with ranks $m=5$ and $m=6$ have an ELPD within one standard deviation of the full rank model, and thus can be viewed as statistically indistinguishable from the model in which there are no constraints on the residual covariance of the outcomes (see Appendix \ref{sec:alcohol_additional}, Table \ref{tab:elpd}). For the remainder of the analysis, we proceed with the rank-5 model as the smallest model which can explain the correlations in the data (see Appendix \ref{sec:alcohol_additional}, Figure \ref{fig:nhanes_heat} for a heatmap of the inferred $\Gamma$).

Next, we use observed covariates to compute benchmark values for the sensitivity value in the $\Lambda$-parameterization (Section \ref{sec:cali_binary_t}).  We find that age, gender, and education are significant predictors for almost every outcome (Appendix \ref{sec:alcohol_additional}, Table \ref{tab:obs_partial_rsq}) and are also all significantly correlated with the propensity for light drinking. We then compute how much the predicted odds of light drinking would change if any of the observed covariates---age, gender, or education---were omitted from the analysis. The empirical 95\% quantile for $\text{Odds}(X)/\text{Odds}(X_{-age})$ = 3.5, which means that for 95\% of the observed units, the predicted odds of light drinking change by a multiplicative factor between $1/3.5$ and $3.5$ when adding age into a treatment model that already included gender and education.  Similarly, we find that the 95\% empirical quantile of $\text{Odds(X)}/\text{Odds}(X_{-gender})$ and $\text{Odds}(X)/\text{Odds}(X_{-education}) $ are both $1.5$.

\begin{figure}	
	\centering
	% \begin{subfigure}[t]{0.75\textwidth}
	% 	\centering
		\includegraphics[width=0.8\textwidth]{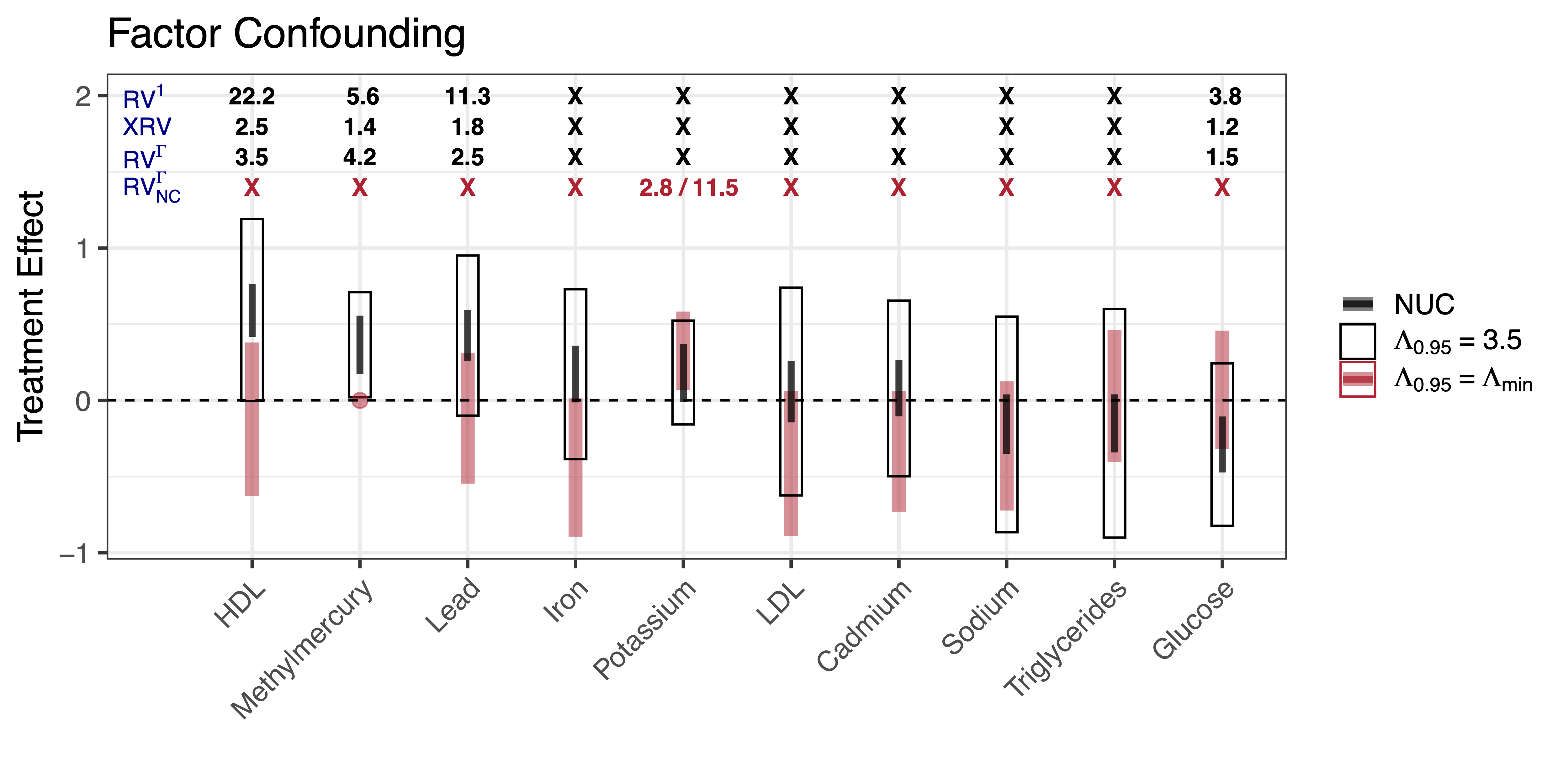}
	% 	\caption{\label{fig:intervals_no_single_outcome_confounders}}	
	% \end{subfigure}
%  \begin{subfigure}[t]{0.75\textwidth}
% 	    \centering
% 	    \includegraphics[width=\textwidth]{figs/effect_intervals_2_odds_m5_null_r2y1_new.pdf}
% 	    \caption{\label{fig:intervals_r2y1}}
%     \end{subfigure}
	 \caption{95\% posterior credible intervals for the causal effects of light drinking on each outcome under NUC (black) and under factor confounding with $\Lambda_{0.95} \leq 3.5$ (black box).  $\Lambda_{0.95}=3.5$ matches the magnitude of the multiplicative change in odds of light drinking when adding age into a propensity model which includes gender and education. In red, we plot the 95\% credible intervals when methylmercury is assumed to be a null control outcome and $\Lambda_{0.95} = \Lambda_{min}$, the value needed to nullify methylmercury under factor confounding (red intervals).  Numbers in black indicate
	 different robustness values, converted to the $\Lambda_{0.95}$-parameterization, including the single outcome robustness values ($RV^1$), extreme robustness ($XRV$), robustness under factor confounding ($RV^\Gamma$) and the combined robustness value with methylmercury as a null control, $RV^\Gamma_{NC} \geq \Lambda_{min, \mathcal{C}}$ (in red). 
	 %$\Lambda^{RV}_{0.95}$ for the outcomes whose credible intervals did not overlap zero under NUC and numbers in red indicate the combined robustness, $\Lambda^{RV}_{0.95, \mathcal{C}} \geq \Lambda_{min}$ after incorporating the null control assumption.  
    Listed values are conservatively reported as the lower endpoint of the 95\% posterior credible intervals of the robustness values. 
% 	 b) Posterior intervals when assuming $R^2_{a'Y \sim U | T, X} = 1$ for methylmercury, so that all residual outcome variance in the null control is due to confounders (predominantly single outcome confounding).  In this setting, the value needed to nullify mercury drops to $\Lambda_{min}=1.4$.  The null control assumption is considerably weaker in this case, and thus has less influence on non-null outcomes.
	 \label{fig:interval_plots}}
\end{figure}

In Figure \ref{fig:interval_plots}, we plot the 95\% posterior intervals for the causal effects on each of the ten outcomes under the no unobserved confounding assumption as black lines. We find that HDL cholesterol, lead, methylmercury, and potassium are positively associated with light drinking and glucose is negatively associated with light drinking.  The 95\% credible intervals for all other outcomes include zero. We also include the worst-case bounds for each outcome assuming $\Lambda_{0.95} = 3.5$ (black rectangles), which matches the benchmark value computed using age, the observed confounder with the strongest relationship to the treatment. While each marginal interval except methylmercury includes zero under this assumption, it is \emph{not} true that all of those outcomes could simultaneously be zero at $\Lambda_{0.95}=3.5$, since the worst-case biases for each outcome are achieved with different values of the sensitivity vector, $\rho$.  

We also include four different robustness values, converted to the  $\Lambda_{0.95}$-parameterization, above each interval. We include the single outcome robustness values \citep[$RV^{1}$,][]{cinelli2020making}, extreme robustness values  \citep[$XRV$,][]{cinelli2022iv}, robustness under factor confounding ($RV^{\Gamma}$) and the combined robustness under factor confounding with methylmercury as a null control ($RV^{\Gamma}_{c=2}$). Since there is estimation uncertainty for these quantities, we conservatively report the lower endpoint of the 95\% posterior interval for each robustness value.  We find that under the factor confounding assumption, methylmercury levels are the most robust to confounding ($RV^{\Gamma}=4.2$) followed by HDL ($RV^{\Gamma}=3.5$).  These robustness values correspond to multiplicative changes in the odds of light drinking which match or exceed the change observed when adding age into a model that already includes gender and education.  In this analysis, the single outcome robustness values, $RV^{1}$, are all much larger than $RV^\Gamma$, and likely overstate the robustness of the effects under NUC.  In contrast, the inferred factor model implies that for many outcomes, a relatively large fraction of outcome variance may be due to unobserved confounding.  As such, for biases large enough to change the sign of the causal effects, $R^2_{T\sim U |X} \ll \rsqytx$ for most outcomes. $XRV$ are by definition the most conservative possible robustness values and thus always less than $RV^{\Gamma}$.

% (Proposition \ref{prop:lambda}). 

% The marginal robustness values for each of these estimates under factor confounding are reported in the 

% As a conservative measure of robustness to confounding, we report the lower 5\% quantile of the posterior distribution of the robustness value for each outcome and print the values in black above the corresponding intervals.  

Despite the apparent robustness of these effects,  we suspect that methylmercury levels are primarily tied to the consumption of fish and note that mercury is not found in alcoholic beverages.  Since there is no other known credible mechanism for light drinking directly influencing mercury levels, methylmercury makes an ideal null control outcome. We then evaluate whether correcting the inferred bias in effect estimates for mercury also explains away the apparent effects for the other outcomes. We plot 95\% posterior credible intervals for all effects after incorporating methylmercury as a null control, assuming no additional confounding beyond the smallest amount required to explain away methylmercury's association.  Stated differently, these plots show the posterior distribution of $a'\teNaive - a' \Gamma \Gamma_{\mathcal{C}}^{\dagger} \teNaiveNull$ which corresponds to the midpoint of the causal effect ignorance regions for outcome $a'Y$ given null controls $\mathcal{C}$, or equivalently,  the identifiable effect estimate under the assumption that $\Lambda_{0.95} =\Lambda_{min, \mathcal{C}}$ (Theorem \ref{thm:ignorance-region-gaussian-wnc,multi-y}).  If the posterior interval for $a'\teNaive - a' \Gamma \Gamma_{\mathcal{C}}^{\dagger} \teNaiveNull$ includes 0, then apparently the null control assumption is enough to explain away the significance of $a'Y$ without introducing additional unobserved confounding.  If it does not include zero, then in red we report the combined robustness, $RV^{\Gamma}_{c=2}$, corresponding to the magnitude of confounding needed to nullify the effect of light drinking on mercury \emph{and} nullify the effect on the corresponding outcome (Theorem \ref{corollary:rv_increase}, right of slash).  We separately report the additional amount needed to nullify the effect beyond what is needed to nullify methylmercury (left of slash). 

When methylmercury is a null control,  $\Lambda_{min, \mathcal{C}} = 4.2$, mercury's robustness value. The 95\% posterior credible intervals of $a'\teNaive - a' \Gamma \Gamma_{\mathcal{C}}^{\dagger} \teNaiveNull$ for HDL, lead, and glucose all include zero after incorporating the null control constraint, meaning that factor confounding and the null control assumption is enough to explain away the causal effects for these outcomes, assuming no additional confounding beyond the minimum needed to explain away the effect on mercury. In contrast, the 95\% credible interval for the causal effect on potassium, which included zero under NUC, moves away from zero since potassium levels are negatively correlated with mercury levels.  On top of the confounders needed to nullify methylmercury, we would need additional confounders that were associated with potassium to change the odds of treatment by as much as about 2.8 (left of slash).  In total, we would need unobserved confounders to change the odds of treatment by a multiplicative factor of as much as $4.2 \times 2.8=11.5$ to nullify both methylmercury and potassium (the combined robustness value, right of slash). When factor confounding holds, we conclude that there is evidence that light alcohol consumption has a positive causal effect on potassium levels, but no apparent effect on HDL, lead, or glucose levels.  In summary, under factor confounding, a relatively strong association between confounders and treatment is needed to explain the bias in mercury, which further implies a large bias adjustment for potentially non-null outcomes.  

Crucially, the factor confounding assumption (Assumption \ref{asm:fact_conf}) can be violated if, for example, some unmeasured environmental confounders influenced both a single outcome and the propensity for light drinking, but not any other outcomes.  In such a case, we could further explore relaxations of factor confounding by manually calibrating $\rsqytx$ to values that are larger than the inferred values.  We discuss such strategies and provide additional theory about generalizations in Appendix \ref{sec:relaxing}. In Appendix \ref{sec:alcohol_additional} we provide several additional results under relaxations of the factor confounding assumption for this example, focused in particular on the potential for single outcome confounding for methylmercury.  At one extreme, when we fix $\rsqytx=1$ for methylmercury, we assume the strongest possible association between confounders and mercury levels. When this holds, the majority of unobserved confounding for methylmercury is uncorrelated with other outcomes, and thus the null control assumption has a much smaller impact on the conclusions for other outcomes (see Appendix \ref{sec:alcohol_additional}, Figure \ref{fig:interval_plots2}). In full generality, we note that each of the $q \times m$ entries of $\Gamma$ can always be specified manually by a practitioner, although in practice it is likely to be very difficult to rigorously justify any such choice without starting from higher level assumptions like those proposed in this work.

\section{Discussion}

In this paper, we propose a sensitivity analysis for characterizing the range of potential biases that can arise in observational analyses with multivariate outcomes. Unlike previous work on observational causal inference with multivariate outcomes, which typically require stronger assumptions for causal identification, we explore the range of causal effects that are compatible with the observed data under different untestable assumptions about the strength of unobserved confounding. We show precisely how the bias varies by outcome and depends on the inferred residual covariance in the outcome model.  When appropriate, we show how assumptions about factor model identifiability can be used to provide stronger results about the robustness of effects. We then characterize how null control outcomes influence both the partial identification region and robustness of effects. 

There are several extensions and generalizations that are worth considering. Importantly, in this work, we focus primarily on modeling under a factor model structure, although extensions for non-continuous outcomes and more complex structures could be developed, perhaps by leveraging the copula decomposition proposed by \citet{zheng2021copula} or by making use of generalized latent variable models \citep{skrondal2004generalized}. Also, in the simulation and NHANES example, we used Bayesian inference, and there is room for a more in-depth exploration of the effects of different prior distributions on partially identified parameters \citep[]{gustafson2015bayesian, zheng2021bayesian}.  

% As one example, the null controls assumption can be viewed as degenerate point mass prior distributions on certain causal effects, and this can easily be relaxed to allow for informative, but nondegenerate prior distributions on causal effects.  Relatedly, we can incorporate assumptions about sparse effects in our framework, e.g. as in \citet{wang2017confounder}, by employing sparse Bayesian priors like the horseshoe \citep{carvalho2009handling} and studying the posterior implications for $\rsqtx$ under these priors.

Finally, this work builds on closely related work on sensitivity analyses for multi-treatment causal analyses \citep{zheng2021copula}.  A natural follow-up would be to consider the identification implications for data that involve both multiple simultaneously applied treatments and multiple outcomes. This would further bridge our work and recent works in proximal causal inference \citep{tchetgen2020introduction}. There may be particular connections to the work of \citet{miao2018identifying} who discuss conditions under which the average treatment effects can be nonparametrically identified with a single null control treatment and a null control outcome, via a double null controls design \citep[see also][]{miao2018confounding, shi2020multiply}. 

% We leave any deeper exploration of these connections for future work.  

% \clearpage
\singlespacing
\bibliographystyle{chicago}
\bibliography{references}

\newpage
\appendix
\section{Theory}

\subsection{Proof of Theorem \ref{thm:ignorance_region_general,multi-y} and Corollary \ref{cor:ignorance_region_global,multi-y}}

\subsubsection*{Proof of Theorem \ref{thm:ignorance_region_general,multi-y}}

\noindent \textbf{Theorem \ref{thm:ignorance_region_general,multi-y}.} 
\begin{itshape}
Assume model \eqref{eqn:u}-\eqref{eqn:epsilon_y} with $\psi_T$ defined by \eqref{eqn:conditional_u_mean}-\eqref{eqn:conditional_u_cov}, and Assumptions \ref{asm:latent-unconfoundedness-scm}-\ref{asm:homoscedasticity}, and let $\sigma^2_{a'Y} := Var(a'Y \mid T=t, X=x)$. The partial fraction of outcome variance explained by confounders is $\rsqytx := ||a'\Gamma||/\sigma^2_{a'Y}$. The confounding bias of $\text{PATE}_{a,t_1,t_2}$, $\text{Bias}_{a,t_1,t_2}$ is equal to $\frac{a' \Gamma \SigmaU^{-1/2}\rho}{\sigma}(t_1 - t_2)$ and it is bounded by
    \begin{equation}
    \label{eqn:worst-case-bias-general,multi-y}
        \text{Bias}_{a,t_1,t_2}^2 
        \, \leq \, 
        \frac{(t_1-t_2)^2}{\sigma^2} \left(\frac{\rsqtx}{1 - \rsqtx}
        \right)\parallel a' \Gamma\parallel_2^2 
        \, \leq \,
        \frac{(t_1-t_2)^2}{\sigma^2} \left(\frac{\rsqtx}{1 - \rsqtx}
        \right) \sigma^2_{a'Y}.
    \end{equation}
The first bound is achieved when $\rho$ is collinear with $a'\Gamma$ and the second bound is achieved when $\rsqytx = 1$. 
\end{itshape}

\begin{proof}
Under model \eqref{eqn:u}-\eqref{eqn:epsilon_y}, we have the following equations:
\begin{align}
    E[Y\mid T=t,X=x] &= g(t,x) + \Gamma\Sigma_{u\mid t, x}^{-1/2}\mu_{u\mid t,x}, \\ 
    E[Y\mid do(T=t), X=x] &= g(t,x) + \Gamma\Sigma_{u\mid t, x}^{-1/2}\mu_{u\mid x}.
\end{align}

\noindent Note that with \eqref{eqn:conditional_u_mean}-\eqref{eqn:conditional_u_cov}, $\sigma^2$ and $\Sigma_{u\mid t, x} = \SigmaU$ do not change with $t$ and $x$. So the confounding bias is equal to 
\begin{align}
    \text{Bias}_{a,t_1,t_2} &= (E[a'Y \mid t_1] - E[a'Y \mid t_2]) - (E[a'Y \mid do(t_1)] - E[a'Y \mid do(t_2)]) \\
    &= E_X[a'(\Gamma\SigmaU^{-1/2}\mu_{u\mid t_1,x} - \Gamma\SigmaU^{-1/2}\mu_{u\mid t_2,x})] \\
    &= \frac{a'\Gamma \SigmaU^{-1/2} \rho}{\sigma}(t_1 - t_2),
\end{align}
where $\SigmaU = I_m - \rho\rho^{\prime}$.

% The sensitivity parameter $\beta$ can be reparameterized in terms of length $d^\beta$ and direction $u^\beta$:
% \begin{equation}
%     \beta = d^\beta u^\beta,
% \end{equation}
% where $d^\beta = \sigma \sqrt{R_{T \sim U\mid X}^2}$ and $u^\beta \in \mathcal{C}^{m-1}$ is a m-dimensional unit vector.
We can write the eigendecomposition of matrix $I_m - \rho\rho^{\prime}$ as
\begin{align}
    I_m - \rho\rho^{\prime} &= U
    \begin{bmatrix}
    1-\|\rho\|^2_2 & & &\\
    & 1 & &\\
    & & \ddots & \\
    & & & 1 \\
    \end{bmatrix}
    U^T,
\end{align}
where $U$ is an orthogonal matrix with the first column as $\frac{\rho}{\|\rho\|_2}$. Thus, $\text{Bias}_{a,t_1,t_2}$ can be simplified as
\begin{align}
    |\text{Bias}_{a,t_1,t_2}| &= |\frac{1}{\sigma} a' \Gamma (I_m - \rho\rho^{\prime})^{-1/2}\rho(t_1 - t_2)| \\
    &= \frac{1}{\sigma\|\rho\|_2} \sqrt{\frac{R_{T \sim U\mid X}^2}{1 - R_{T \sim U \mid X}^2}} |a' \Gamma \rho(t_1 - t_2)|  \label{eqn:bias_deltat} \\
    &\leq \frac{1}{\sigma} \sqrt{\frac{R_{T \sim U \mid X}^2}{1 - R_{T \sim U \mid X}^2}} \parallel a'\Gamma(t_1 - t_2) \parallel_2,
\end{align}
where the bound is reached when $\rho$ is collinear with $ \Gamma'a$. And note that $Var(a'Y \mid T=t, X=x) - a'\Gamma\Gamma'a = a'\Delta a \geq 0$, the inequalities are proved.
\end{proof}

\subsubsection*{Proof of Corollary \ref{cor:ignorance_region_global,multi-y}}

\noindent \textbf{Corollary \ref{cor:ignorance_region_global,multi-y}.}
\begin{itshape}
Let $d_1$ be the largest singular value of $\Gamma$. For all unit vectors $a$, the confounding bias, $\text{Bias}_{a,t_1,t_2}$, is bounded by
\begin{equation}
    \text{Bias}_{a,t_1,t_2}^2 \leq \frac{(t_1-t_2)^2}{\sigma^2} \frac{R_{T \sim U\mid X}^2}{1 - R_{T \sim U \mid X}^2}d_1^2,
\end{equation}
with equality when $a = u_1^{\Gamma}$, the first left singular vector of $\Gamma$, and when $\rho$ is collinear with $v_1^{\Gamma}$, the first right singular vector of $\Gamma$. There is no confounding bias for the causal effect estimates of outcome $a'Y$ when $a \in Null(\Gamma')$.
\end{itshape}
\begin{proof}
From Equation \eqref{eqn:bias_deltat}, the confounding bias is 
\begin{equation}
    \text{Bias}_{a,t_1,t_2}  = \frac{t_1 - t_2}{\sigma\|\rho\|_2} \sqrt{\frac{R_{T \sim U\mid X}^2}{1 - R_{T \sim U \mid X}^2}}a' \Gamma \rho.
\end{equation}
According to the property of Rayleigh quotient, we have $(a'\Gamma \rho)^2 \leq \| a' \Gamma\|_2^2 \|\rho\|_2^2\leq d_1^2\|\rho\|_2^2$. $a' \Gamma \rho/\|\rho\|_2$ reaches its maximum, $d_1$, the largest singular value of $\Gamma$, when $a = u_1^{\Gamma}$, the first left singular vector of $\Gamma$, and $\rho$ is collinear with $v_1^{\Gamma}$, the first right singular vector of $\Gamma$. Thus, 
\begin{equation}
    \text{Bias}_{a,t_1,t_2}^2 \leq \frac{d_1^2}{\sigma^2} \frac{R_{T \sim U\mid X}^2}{1 - R_{T \sim U \mid X}^2}(t_1-t_2)^2.
\end{equation}
When $a \in Null(\Gamma')$, we have $\text{Bias}_{a,t_1,t_2}=0$. 
\end{proof}

\subsection{Proof of Proposition \ref{prop:cali_wnco}, Theorem \ref{thm:ignorance-region-gaussian-wnc,multi-y} and Theorem \ref{corollary:rv_increase}}

\subsubsection*{Proof of Proposition \ref{prop:cali_wnco}}

\textbf{Proposition \ref{prop:cali_wnco}.}
\begin{itshape}
Assume model \eqref{eqn:u}-\eqref{eqn:epsilon_y} with sensitivity parameterization \eqref{eqn:conditional_u_mean}-\eqref{eqn:conditional_u_cov} and Assumptions \ref{asm:latent-unconfoundedness-scm}-\ref{asm:homoscedasticity}. Further, suppose there are $c$ null control outcomes, $Y_j$, such that $\tau_j = 0$ for $j \in \mathcal{C}$. Then, $\teNaiveNull$ must be in the column space of $\Gamma_{\mathcal{C}}$. In addition, the fraction of variation in the treatment due to the confounding is lower bounded by
\begin{equation}
    \rsqtx \geq R_{\text{min}, \mathcal{C}}^2 := \frac{\sigma^2 \parallel 
    \Gamma_{\mathcal{C}}^{\dagger} \teNaiveNull \parallel_2^2}{(t_1-t_2)^2+\sigma^2 \parallel 
    \Gamma_{\mathcal{C}}^{\dagger} \teNaiveNull \parallel_2^2},
\end{equation}
where $\Gamma_{\mathcal{C}}^{\dagger}$ denotes the pseudoinverse of $\Gamma_{\mathcal{C}}$. $R_{\text{min}, \mathcal{C}}^2$ is identifiable under factor confounding (Assumption \ref{asm:fact_conf}).
\end{itshape}
\begin{proof}
Assume there are $c$ null control outcomes, satisfying 
\begin{equation}
\label{eqn:nc_constraint,multi-y,proof}
   \teNaiveNull = \frac{t_1 - t_2}{\sigma\sqrt{1 - \rsqtx}} \Gamma_{\mathcal{C}} \rho,
\end{equation}
The solution of the above equation exists if and only if $ Q_{\Gamma_{\mathcal{C}}} \teNaiveNull = \teNaiveNull$ holds, where $Q_{\Gamma_{\mathcal{C}}} =\Gamma_{\mathcal{C}}\Gamma_{\mathcal{C}}^{\dagger}$ is the projection matrix onto the column space of $\Gamma_{\mathcal{C}}$. Under this condition, all solutions to Equation \eqref{eqn:nc_constraint,multi-y,proof} can be written as 
\begin{equation}
    \rho = \frac{\sigma}{t_1-t_2} \sqrt{1 - \rsqtx} \ 
     \Gamma_{\mathcal{C}}^{\dagger} \teNaiveNull
    + (I_m -  \Gamma_{\mathcal{C}}^{\dagger} \Gamma_{\mathcal{C}}) w.
\end{equation}
Since $\rho'\rho = \rsqtx$, $w$ can be any $m \times 1$ vector satisfying
\begin{equation}
    \parallel (I_m -  \Gamma_{\mathcal{C}}^{\dagger} \Gamma_{\mathcal{C}}) w \parallel_2^2 \,= \rsqtx - \frac{\sigma^2(1 - \rsqtx)}{(t_1-t_2)^2}\parallel \Gamma_{\mathcal{C}}^{\dagger} \teNaiveNull \parallel_2^2.
\end{equation}
In addition, since $\parallel (I_m -  \Gamma_{\mathcal{C}}^{\dagger} \Gamma_{\mathcal{C}}) w \parallel_2^2 \,\geq 0$, we prove the result 
\begin{equation}
    \rsqtx \geq R^2_{min, \mathcal{C}} := \frac{\sigma^2 \parallel 
    \Gamma_{\mathcal{C}}^{\dagger} \teNaiveNull \parallel_2^2}{(t_1-t_2)^2 + \sigma^2 \parallel 
    \Gamma_{\mathcal{C}}^{\dagger} \teNaiveNull \parallel_2^2}.
\end{equation}

Under factor confounding, $\Gamma$ is identified up to rotations from the right. Since $\parallel \Gamma_{\mathcal{C}}^{\dagger} \teNaiveNull \parallel_2^2$ is invariant under rotations of $\Gamma$, $R^2_{min, \mathcal{C}}$ is identified.
\end{proof}

\subsubsection*{Proof of Theorem \ref{thm:ignorance-region-gaussian-wnc,multi-y}}

\textbf{Theorem \ref{thm:ignorance-region-gaussian-wnc,multi-y}.}
\begin{itshape} 
Under the assumptions established in Proposition \ref{prop:cali_wnco}, for any value of $\rsqtx \geq R^2_{min, \mathcal{C}}$, the confounding bias for the treatment effect on outcome $a'Y$ is in the interval
\begin{equation}
    \text{Bias}_{a,t_1,t_2} \in 
    \left[a' \Gamma \Gamma_{\mathcal{C}}^{\dagger} \teNaiveNull
    \; \pm \;
    \parallel
    a'\Gamma P_{\Gamma_{\mathcal{C}}}^{\perp} 
    \parallel_2
    \sqrt{
    \frac{(t_1-t_2)^2}{\sigma^2}\left(
    \frac{\rsqtx}{1 - \rsqtx} - 
    \frac{R^2_{min, \mathcal{C}}}{1 - R^2_{min, \mathcal{C}}} 
    \right)}\,
\right],
\end{equation}
where $P_{\Gamma_{\mathcal{C}}}^{\perp} = I_m - \Gamma_{\mathcal{C}}^{\dagger} \Gamma_{\mathcal{C}}$ is the $m \times m$ projection matrix onto the space orthogonal to the row space of $\Gamma_{\mathcal{C}}$. Under Assumption \ref{asm:fact_conf}, $\rsqtx$ is the only unidentifiable parameter.
\end{itshape}

\begin{proof}
According to Proposition \ref{prop:cali_wnco}, the confounding bias of outcome $a'Y$ is 
\begin{align}
\text{Bias}_{a,t_1,t_2} &= \frac{t_1 - t_2}{\sigma\sqrt{1 - \rsqtx}} a'\Gamma\rho \\
&= a'\Gamma  \Gamma_{\mathcal{C}}^{\dagger} \teNaiveNull + \frac{t_1-t_2}{\sigma\sqrt{1 - \rsqtx}} a'\Gamma (I_m -  \Gamma_{\mathcal{C}}^{\dagger} \Gamma_{\mathcal{C}})^2w.
\end{align}
Note that we have 
\begin{equation}
    |a'\Gamma (I_m -  \Gamma_{\mathcal{C}}^{\dagger} \Gamma_{\mathcal{C}})^2w| \leq \parallel a'\Gamma P_{\Gamma_{\mathcal{C}}}^{\perp} \parallel_2 \sqrt{\rsqtx - \frac{\sigma^2(1 - \rsqtx)}{(t_1-t_2)^2}\parallel \Gamma_{\mathcal{C}}^{\dagger} \teNaiveNull \parallel_2^2},
\end{equation}
with $P_{\Gamma_{\mathcal{C}}}^{\perp}:= I_m - \Gamma_{\mathcal{C}}^{\dagger} \Gamma_{\mathcal{C}}$, and the bound is achieved when $P_{\Gamma_{\mathcal{C}}}^{\perp} w$ is collinear with $P_{\Gamma_{\mathcal{C}}}^{\perp}\Gamma'a$. Combine this with the definition of $R^2_{min, \mathcal{C}}$, we prove that the confounding bias of $a'Y$ is in the interval
\begin{equation}
    a' \Gamma \Gamma_{\mathcal{C}}^{\dagger} \teNaiveNull
     \; \pm \;
    \parallel
    a'\Gamma P_{\Gamma_{\mathcal{C}}}^{\perp} 
    \parallel_2 
    \sqrt{ \frac{(t_1-t_2)^2}{\sigma^2}\left(
    \frac{\rsqtx}{1 - \rsqtx} - 
    \frac{R^2_{min, \mathcal{C}}}{1 - R^2_{min, \mathcal{C}}} 
    \right)}.
\end{equation}

From Proposition \ref{prop:cali_wnco}, $R_{min, \mathcal{C}}^2$ is identifiable under Assumption \ref{asm:fact_conf}, and note that $\Gamma \Gamma_{\mathcal{C}}^{\dagger}$ and $\parallel a'\Gamma P_{\Gamma_{\mathcal{C}}}^{\perp} \parallel_2$ are invariant to the rotations of $\Gamma$. Hence, $\rsqtx$ is the only unidentified parameter in this interval.
\end{proof}

\subsubsection*{Proof of Theorem \ref{corollary:rv_increase}}

\textbf{Theorem \ref{corollary:rv_increase}.}
\begin{itshape}
Let $\teNaive$ denote the vector of PATEs for all outcomes and $\teNaiveNull$ be the vector of PATEs for null control outcomes under NUC. Under the assumptions established in Proposition \ref{prop:cali_wnco}, the factor confounding robustness value for outcome $a'Y$ is
\begin{equation}
RV^{\Gamma}_{a} = \frac{\omega}{1+\omega},
\end{equation}
\noindent where $\omega = \frac{\sigmat^2}{(t_1-t_2)^2}\frac{(a'\teNaive)^2}{ \parallel a'\Gamma \parallel_2^2}$ . The combined robustness value for outcome $a'Y$ given null controls $\mathcal{C}$ is
\begin{equation}
RV_{a,\mathcal{C}}^\Gamma = \frac{w_{\mathcal{C}}}{1 + w_\mathcal{C}}\geq max(R^2_{min, \mathcal{C}}, RV^{\Gamma}_{a}),
\end{equation}
where $w_{\mathcal{C}} := \frac{\sigma^2}{(t_1-t_2)^2} \left[\frac{(a'(\teNaive - \Gamma\Gamma^{\dagger}_{\mathcal{C}}\teNaiveNull))^2}{\parallel a'\Gamma P_{\Gamma_{\mathcal{C}}}^{\perp} \parallel^2_2} + \parallel \Gamma_{\mathcal{C}}^{\dagger}\teNaiveNull\parallel^2_2\right]$.
\end{itshape}

\begin{proof}
It is possible to consider the robustness value only when $\text{Nullity}(\Gamma_{\mathcal{C}}) > 0$ and $a'\Gamma$ is not in the row space of $\Gamma_{\mathcal{C}}$. Based on Equation \eqref{eqn:ignorance-region-gaussian-wnc,multi-y}, the bounds of the ignorance region for $\text{PATE}_{a,t_1,t_2}$ can be written as 
\begin{equation}
\label{eqn:ig-bound-nc-gaussian}
    a'\teNaive - a' \Gamma \Gamma_{\mathcal{C}}^{\dagger} \teNaiveNull
    \; \pm \;
    \parallel
    a'\Gamma P_{\Gamma_{\mathcal{C}}}^{\perp} 
    \parallel_2
    \sqrt{
    \frac{(t_1-t_2)^2}{\sigma^2}\left(
    \frac{\rsqtx}{1 - \rsqtx} - 
    \frac{R^2_{min, \mathcal{C}}}{1 -R^2_{min, \mathcal{C}}} 
    \right)}.
\end{equation}
To calculate the minimum amount of confounding needed to change the sign of treatment effect, we can set Equation \eqref{eqn:ig-bound-nc-gaussian} to zero and solve for $\rsqtx$. Therefore, we have 
\begin{align}
    \frac{(a'\teNaive - a' \Gamma \Gamma_{\mathcal{C}}^{\dagger} \teNaiveNull)^2}{\parallel
    a'\Gamma P_{\Gamma_{\mathcal{C}}}^{\perp} 
    \parallel_2^2}
    &= 
    \frac{(t_1-t_2)^2 RV_{a,\mathcal{C}}^\Gamma}{\sigma^2(1 - RV_{a,\mathcal{C}}^\Gamma)} - \parallel \Gamma_{\mathcal{C}}^{\dagger} \teNaiveNull \parallel_2^2 \\
    \Rightarrow \frac{RV_{a,\mathcal{C}}^\Gamma}{1 - RV_{a,\mathcal{C}}^\Gamma}
    &= \frac{\sigma^2}{(t_1-t_2)^2} \left[\frac{(a'(\teNaive - \Gamma\Gamma^{\dagger}_{\mathcal{C}}\teNaiveNull))^2}{\parallel a'\Gamma P_{\Gamma_{\mathcal{C}}}^{\perp} \parallel^2_2} + \parallel \Gamma_{\mathcal{C}}^{\dagger}\teNaiveNull\parallel^2_2\right]
    \label{eqn:rv-nc-gaussian-mediate} \\
    \Rightarrow RV_{a,\mathcal{C}}^\Gamma &= \frac{w_{\mathcal{C}}}{1 + w_\mathcal{C}},
\end{align}
where $w_{\mathcal{C}}$ denotes the right-hand side of Equation \eqref{eqn:rv-nc-gaussian-mediate}. $RV_{a,\mathcal{C}}^\Gamma$ must be greater than $R^2_{min, \mathcal{C}}$ and $RV^{\Gamma}_{a}$ according to the definition in \eqref{def:combined-rv}. We can derive $RV^\Gamma_a$ in the same way and it is equal to $RV_{a,\mathcal{C}}^\Gamma$ where $\mathcal{C}$ is empty.
\end{proof}

\subsection{Proof of Proposition \ref{prop:lambda}}

\textbf{Proposition \ref{prop:lambda}.}
\begin{itshape}
Assume $U$ is conditionally Gaussian with mean $\mu_{u\mid t,x}$ and covariance $\SigmaU$ as given in Equations \eqref{eqn:conditional_u_mean} and \eqref{eqn:conditional_u_cov}.  Further, denote $e(x) = P(T=1\mid X=x)$ and $e_0(x, U) = P(T=1 \mid X=x, U)$.  Then, for any $\rho$ we have $\lambda(X=x, U) = \frac{e_0(x, U)/(1-e_0(x, U))}{e(x)/(1-e(x))} = \text{exp}(V_x)$, where $V_x = (2I_x-1)Z$, $I_x\sim \text{Ber}(e(x)), Z \sim N(\mu_\lambda, \sigma^2_\lambda )$ with $\mu_\lambda = \frac{1}{2\sigma^2}\frac{\rsqtx}{1-\rsqtx}$ and $\sigma^2_\lambda = \frac{1}{\sigma^2}\frac{\rsqtx}{1-\rsqtx}$.
\end{itshape}

\begin{proof}
According to Bayes' rule and the conditional distribution of $U$ given $T$ and $X$, we can write the odds of treatment after accounting for unmeasured confounders as
\begin{align}
    \frac{e_0(x, u)}{1-e_0(x, u)} &= \frac{f(u\mid T=1, x)e(x)}{f(u\mid T=0, x)(1 - e(x))} \\
    \begin{split}
    &= \frac{e(x)}{1-e(x)}\text{exp}\Big[\frac{1}{2}(\mu_{u\mid T=0,x}'\Sigma_{u\mid t,x}^{-1}\mu_{u\mid T=0,x} - \mu_{u\mid T=1,x}'\Sigma_{u\mid t,x}^{-1}\mu_{u\mid T=1,x}) \\
    &\quad +(\mu_{u\mid T=1,x}' - \mu_{u\mid T=0,x}')\Sigma_{u\mid t,x}^{-1}u \Big].    
    \end{split}
    \label{eqn:true_odds}
\end{align}
Then from Equations \eqref{eqn:conditional_u_mean} and \eqref{eqn:conditional_u_cov}, we have
\begin{align}
    (\mu_{u\mid T=1,x}' - \mu_{u\mid T=0,x}')\Sigma_{u\mid t,x}^{-1} &= \frac{\rho'}{\sigma}\left(I_m - \rho\rho'\right)^{-1} \\
    &= \frac{\rho'}{\sigma(1 - \parallel \rho \parallel_2^2)}.
\end{align}
Note that $\mu_{t\mid x} = e(x)$, then from Equation \eqref{eqn:true_odds}, the odds ratio given $x$ can be written as 
\begin{align}
    \lambda(x, U) &\overset{d}{=} \text{exp}\left[\frac{2e(x) - 1}{2\sigma^2}\rho'\Sigma_{u\mid t,x}^{-1}\rho + \frac{\rho'U_x}{\sigma - \sigma\parallel \rho \parallel_2^2}\right] \\
    &= \text{exp}\left[\frac{(2e(x) - 1)\parallel \rho \parallel_2^2}{2\sigma^2(1 - \parallel \rho \parallel_2^2)} + \frac{\rho'U_x}{\sigma - \sigma\parallel \rho \parallel_2^2}\right], 
\end{align}
where $U_x \sim f(u\mid x)$. Denote the indicator function $\mathbbm{1}\{T=1\mid x\}$ as $I_x \sim \text{Ber}(e(x))$. Then $\rho'U_x$ can be written as a two-component mixture of $(1 - I_x)V_0 + I_x V_1$, where $V_0 = \rho'U\mid T=0,x$ and $V_1 = \rho'U\mid T=1,x$. From the distribution of $U$ given $T$ and $X$, we have $V_0 \sim N(\frac{- \parallel \rho \parallel_2^2 e(x)}{\sigma}, \| \rho \|_2^2(1 -  \| \rho \|_2^2))$ and $V_1 \sim N(\frac{\parallel \rho \parallel_2^2 (1-e(x))}{\sigma}, \| \rho \|_2^2(1 -  \| \rho \|_2^2))$. Therefore, the odds ratio given $x$ has the following distribution 
\begin{align}
    \lambda(x, U) &\overset{d}{=} \text{exp}\left[\frac{(2e(x) - 1)\parallel \rho \parallel_2^2}{2\sigma^2(1 - \parallel \rho \parallel_2^2)} + \frac{(1 - I_x)V_0 + I_x V_1}{\sigma - \sigma\parallel \rho \parallel_2^2}\right] \\
    &\overset{d}{=} \text{exp}[(2I_x-1)Z],
\end{align}
where $Z \sim N(\mu_\lambda, \sigma^2_\lambda )$ with $\mu_\lambda = \frac{1}{2\sigma^2}\frac{\rsqtx}{1-\rsqtx}$ and $\sigma^2_\lambda = \frac{1}{\sigma^2}\frac{\rsqtx}{1-\rsqtx}$.
\end{proof}

% \subsection{Proof of Proposition \ref{prop:nc_bias_correction}}

% \textbf{Proposition \ref{prop:nc_bias_correction}}
% \begin{itshape}
% Let $\tilde \Gamma = (\hat \Gamma \hat \Gamma' + D_\mathcal{C})^{1/2}$, where $D_\mathcal{C}$ is zero everywhere except for nonnegative entries along the diagonal elements corresponding to the null control outcomes. Then,
% \begin{equation}
% |a'\tilde \Gamma \tilde \Gamma^{\dagger}_{\mathcal{C}} \teNaive_{\mathcal{C}}| \leq |a'\hat \Gamma\hat \Gamma^{\dagger}_{\mathcal{C}} \teNaive_{\mathcal{C}}| 
% \end{equation}
% with equality if and only if $D_\mathcal{C}=0$.
% \end{itshape}

% \proof{
% 1
% }

\pagebreak

\section{Relaxations of Factor Confounding}

\label{sec:relaxing}

We now return to the factor confounding assumption (Assumption \ref{asm:fact_conf}).   Here, we explore the ways in which this assumption might plausibly be violated and characterize the effects that such violations might have on the causal effect bounds.  First, we consider the consequences of violating the assumption that the residual confounding is induced by unmeasured pre-treatment variables (Definition \ref{asm:potential_conf}).  Some function of the latent factors, $U$, may reflect unmeasured post-treatment mediators or even direct dependence amongst the outcomes. In this case, conditional on all unmeasured pre-treatment variables, there would still be residual correlations in the outcomes from these unmeasured post-treatment sources.  Here, we show that regardless of the source of residual correlations in the outcomes, the bias of the causal effect estimate will never exceed the bounds established in Theorem \ref{thm:ignorance_region_general,multi-y}.  

\begin{proposition}
Let $U_1 = AU$, where $A$ is an $r \times m$ semi-orthogonal matrix with $0 \le r \le m$. Assume latent unconfoundedness holds given just $U_1$ and the other assumptions follow Theorem \ref{thm:ignorance_region_general,multi-y}, then we can rewrite Equation \eqref{eqn:epsilon_y} as $Y = g(T, X) + \Gamma_1 \Sigma_{u_1 \mid t,x}^{-1/2} U_1 + \epsilon_1$, with $Cov(\epsilon_1 \mid t,x) = M + \Delta$ for a positive semi-definite matrix $M$, and $\epsilon_1$ independent of $U_1$ conditional on $T$ and $X$. 
Then, the confounding bias for outcome $a'Y$ is still in the intervals defined in Theorem \ref{thm:ignorance_region_general,multi-y}.
%
% \noindent Proof: See appendix.
\label{cor:conservative}

\begin{proof}

% Assume $U=(U_1^T, U_2^T)^T$ without loss of generality. Suppose the dimension of $U_1$ is $m_1$, the dimension of $U_2$ is $m_2$, with $m_1 + m_2 = m$. Denote $\Sigma_{u \mid t}^{-1/2}=(\Sigma_1, \Sigma_2)$, where $\Sigma_1$ is $m$ by $m_1$, $\Sigma_2$ is $m$ by $m_2$. Then Equation \ref{eqn:outcome,multi-y} can be written as 
% \begin{equation}
%     Y = \tau T + \Gamma\Sigma_1 U_1 + \Gamma\Sigma_2 U_2 + \epsilon_{y|t,u} = \tau T + \Gamma_1 \Sigma_{u_1 \mid t}^{-1/2} U_1 + \Gamma_2 \Sigma_{u_2 \mid t}^{-1/2} U_2 + \epsilon_{y|t,u},
% \end{equation}
% where $\Gamma_i = \Gamma\Sigma_i\Sigma_{u_i \mid t}^{-1/2}$ for $i\in\{1,2\}$.

First, we have $\mu_{u_1\mid t,x} = \frac{\rho_1}{\sigma}\left(t-\mu_{t\mid x}\right)$ and $\Sigma_{u_1 \mid t,x} = I_m-\rho_1 \rho_1^{\prime}$,
where $\rho_1= A\rho$. Since $A$ is $r \times m$ semi-orthogonal, we have $\| \rho \|_2 = \| A'\rho_1 \|_2 =\| \rho_1 \|_2$.

Under the assumptions about $U_1$, the conditional expectation of $Y$ can be written as
\begin{equation}
    E[Y\mid T=t,X=x] = g(t,x) + \Gamma_1\Sigma_{u_1\mid t, x}^{-1/2}\mu_{u_1\mid t,x} + E[\epsilon_1 \mid t,x].
\end{equation}
And the X-specific average causal effect is equal to 
\begin{equation}
    E[Y\mid do(T=t), X=x] = g(t,x) + \Gamma_1\Sigma_{u_1\mid t, x}^{-1/2}\mu_{u_1\mid x} + E[\epsilon_1 \mid t,x].
\end{equation}
Then follow the proof of Theorem \ref{thm:ignorance_region_general,multi-y}, the confounding bias for outcome $a'Y$ is bounded by 
\begin{align}
    |\text{Bias}_{a,t_1,t_2}| &= |\frac{a' \Gamma_1}{\sigma} (I_m - \rho_1\rho_1')^{-1/2} \rho_1(t_1-t_2)|\\
    &\leq \frac{|(t_1-t_2)|}{\sigma} \sqrt{\frac{\rsqtx}{1 - \rsqtx}} \parallel a' \Gamma_1 \parallel_2  \\
    &\leq \frac{|(t_1-t_2)|}{\sigma} \sqrt{\frac{\rsqtx}{1 - \rsqtx}} \parallel a' \Gamma \parallel_2.
\end{align}
The first inequality is due to $\| \rho_1 \|_2^2= \| \rho \|_2^2 = \rsqtx$ and it is achieved when $\rho_1$ is collinear with $\Gamma_1'a$. The second inequality holds since $\Gamma\Gamma' - \Gamma_1\Gamma_1' = M$ is positive semi-definite. Therefore, the confounding bias for outcome $a'Y$ is still in the intervals from Theorem \ref{thm:ignorance_region_general,multi-y}.
\end{proof}
\end{proposition}

\noindent This proposition indicates that violations to Assumption \ref{asm:potential_conf} are relatively innocuous, in that the omitted variable bias for all causal effects still cannot exceed the established bounds.  For methods that address sensitivity to unobserved confounding in mediation analyses, see a closely related approach by \citet{zhang2022interpretable}.

Next, we consider violations of factor confounding that arise because some confounders influence fewer than three outcomes, thus rendering $\Gamma$ non-identifiable, at least for some rows (Theorem \ref{thm:gamma-identifiability}).  Such a violation is of potential concern because it may lead to underestimates of $||a'\Gamma||_2$ and hence underestimates of the potential omitted variable bias. Fortunately, Theorem \ref{thm:ignorance_region_general,multi-y} highlights that even when $\Gamma$ is not identifiable, we can still identify a conservative bound for the omitted variable bias for any outcome by replacing  $\Gamma$ with a square root of the inferred residual covariance, $\Sigma_{y\mid t, x}^{1/2}$. More generally, we might be able to establish less conservative bounds, by calibrating the fraction of each outcome's residual variance explained by potential confounders.  Given $\Gamma$, the fraction of outcome residual variance for outcome $a'Y$ is $R^2_{a'Y \sim U \mid T,X} = \frac{a'\Gamma\Gamma'a}{a'(\Gamma\Gamma' +\Delta)a}$. By making the factor confounding assumption, we avoid having to directly specify $R^2_{a'Y \sim U \mid T,X}$ for each of the $q$ outcomes, but when we suspect factor confounding is violated for outcome $a'Y$, we can make adjustments to $\Gamma$ by reasoning about $R^2_{a'Y \sim U \mid T,X}$.  For these reasons, we view bounds based on factor confounding as a useful reference assumption that must be rigorously justified.  
In the next section, we discuss how to calibrate $R^2_{a'Y\sim U \mid T, X}$ when factor confounding does not hold. In the analysis of Section \ref{sec:nhanes} we demonstrate how to reason about the plausibility of factor confounding in a real data example.

\subsection*{Calibrating outcome variance explained by confounding} 
\label{sec:cal_r2y}

So far, we have focused on settings in which $\Gamma$ is identifiable up to a causal equivalence class (factor confounding), or fixed by assumption (e.g. to $\Sigma^{1/2}_{y\mid t, x}$ which leads to the largest ignorance region for all causal effects given any choice of $\rsqtx$).  As shown in Theorem \ref{thm:gamma-identifiability}, factor confounding requires that each confounder influences multiple outcomes and that the number of confounders, $m$, is not too large relative to the number of outcomes, $q$. Factor confounding makes particular sense in some settings, for instance, when confounding is caused by batch effects \citep{gagnon2013removing} or when we have domain knowledge that specific unmeasured confounders are likely to influence several outcomes. Given that factor confounding rests on these untestable assumptions, we also recommend benchmarking the outcome variance explained by factor confounding, $R^2_{a'Y\sim U \mid T, X} = \frac{\text{Var}(a'Y|T, X) - \text{Var}(a'Y \mid U, T, X)}{\text{Var}(a'Y| T, X)} = \frac{a'\Gamma\Gamma'a}{a'(\Gamma\Gamma' +\Delta)a}$.  

In general, our strategy is to lower bound $\rsqytx$ with values based on the factor confounding assumption and then increase $\rsqytx$ for specific outcomes as needed based on benchmarked values.  Assume that we obtain $\hat \Gamma$ and $\hat \Delta$ by fitting a factor model, and use these to compute
$\frac{a'\hat \Gamma\hat \Gamma'a}{a'(\hat \Gamma\hat \Gamma' +\hat\Delta) a}$.  We can compare this quantity to partial R-squared measures computed from reference covariates.  For a reference covariate $X_j$ and all baseline covariates $X_{-j}$, we can compute the partial R-squared for outcome $a'Y$, $R_{a'Y \sim X_j \mid X_{-j}, T}^2 := \frac{R_{a'Y \sim X, T}^2 - R_{a'Y\sim X_{-j}, T}^2}{1 - R_{a'Y \sim X_{-j}, T}^2}$.  When $\frac{a'\hat \Gamma\hat \Gamma'a}{a'(\hat \Gamma\hat \Gamma' +\hat\Delta) a}$ is large relative to expectations based on these benchmark values, then we might expect bounds for the causal effect on outcome $a'Y$ to be conservative under factor confounding. In cases where $\frac{a'\hat \Gamma\hat \Gamma'a}{a'(\hat \Gamma\hat \Gamma' +\hat\Delta) a}$ is small relative to expectations from benchmark values, we should consider the possibility that factor confounding is violated for outcome $a'Y$ and make appropriate adjustments. For example, there may be confounders of outcome $a'Y$ which are not reflected in the inferred loading matrix $\hat \Gamma$ because they are uncorrelated with other outcomes (``single outcome confounders'').

In this case, we can manually attribute additional residual outcome variance on a per outcome basis by computing bounds on the causal effects using $\Gamma = \text{chol}(\hat \Gamma \hat \Gamma' + D)$ where $D$ is a diagonal matrix such that $0 \preceq D \preceq \hat \Delta$ and $\text{chol}(A)$ denotes the Cholesky factor of matrix $A$. By choosing D appropriately,  we can fix $R^2_{a'Y \sim U \mid T, X}$ to any value between $\frac{a'\hat \Gamma\hat \Gamma'a}{a(\hat \Gamma\hat \Gamma'+\hat\Delta) a}$ and $1$.  The most conservative sensitivity bounds are achieved by using $D = \hat\Delta$ so that $\Gamma\Gamma' = \Sigma_{y\mid t,x}$ and $R^2_{a'Y \sim U \mid T,X} = 1$ for any $a$ (Theorem \ref{thm:ignorance_region_general,multi-y}). As we will illustrate empirically in Section \ref{sec:nhanes}, changing $\rsqytx$ for the null control outcomes also impacts the sensitivity regions for the non-null outcomes.

\subsection*{Heteroscedasticity}

There are two natural ways to extend our sensitivity model to the heteroscedastic case. First, we consider $U$-$X$ interaction in the treatment assignment mechanism \eqref{eqn:treatment_general,multi-y} with $\mu_{u\mid t,x} = \frac{\beta_x}{\sigma^{2}_{t\mid x}}\left(t-\mu_{t\mid x}\right)$ and $\Sigma_{u \mid t,x} = I_m-\frac{\beta_x \beta_x^{\prime}}{\sigma^{2}_{t\mid x}}$, where $\beta_x$ and $\sigma^2_{t\mid x}:= Var(T\mid X=x)$ can vary with $x$. This is motivated by the case where $U$ and $T$ are jointly multivariate Gaussian given $x$. We use the notation $0 \leq \rho_x^2 := \frac{\beta_x^{\prime}\beta_x}{\sigma^2_{t\mid x}} < 1$ to reflect the strength of dependence between the confounders and the treatment and it is equal to $\|\text{Cor}(T, U\mid X=x)\|_2^2$ in the linear model. Second, the outcome model \eqref{eqn:epsilon_y} can be heteroscedastic where $\Gamma_{t,x}$ can vary with $t$ and $x$. Then the confounding bias of $\text{PATE}_{a,t_1,t_2}$ is 
\begin{equation}
    \text{Bias}_{a,t_1,t_2} = E_X\left[\frac{a'\Gamma_{t_1,x}\beta_x}{\sigma_{t\mid x}\sqrt{\sigma^2_{t\mid x} - \beta_x^{\prime}\beta_x}}(t_1 - \mu_{t\mid x}) - \frac{a'\Gamma_{t_2,x}\beta_x}{\sigma_{t\mid x}\sqrt{\sigma^2_{t\mid x} - \beta_x^{\prime}\beta_x}}(t_2 - \mu_{t\mid x})\right].
 \end{equation}
We summarize the bounds of confounding bias for all the outcomes in Theorem \ref{thm:hetero_bound}:

\begin{theorem}
\label{thm:hetero_bound}
Assume model \eqref{eqn:u}-\eqref{eqn:epsilon_y} with $\psi_T$ defined by $\mu_{u\mid t,x} = \frac{\beta_x}{\sigma^{2}_{t\mid x}}\left(t-\mu_{t\mid x}\right)$ and $\Sigma_{u \mid t,x} = I_m-\frac{\beta_x \beta_x^{\prime}}{\sigma^{2}_{t\mid x}}$, and Assumptions \ref{asm:latent-unconfoundedness-scm}-\ref{asm:sutva}. The confounding bias of $\text{PATE}_{a,t_1,t_2}$, $\text{Bias}_{a,t_1,t_2}$ is bounded by
\begin{equation}
|\text{Bias}_{a,t_1,t_2}| \leq E_X\left[\frac{1}{\sigma_{t\mid x}} \sqrt{\frac{\rho_x^2}{1 - \rho_x^2}}(\parallel a' \Gamma_{t_1,x}(t_1 - \mu_{t\mid x})\parallel_2 + \parallel a' \Gamma_{t_2,x}(t_2 - \mu_{t\mid x})\parallel_2) \right], 
\end{equation}
where the bound is achieved when $\beta_x$, $\Gamma_{t_1,x}'a$ and $\Gamma_{t_2,x}'a$ are collinear, $a'\Gamma_{t_1,x}\beta_x(t_1 - \mu_{t\mid x})$ and $a'\Gamma_{t_2,x}\beta_x(t_2 - \mu_{t\mid x})$ have constant opposite signs for all $x$. If we choose the homoscedastic sensitivity parameterization \eqref{eqn:conditional_u_mean}-\eqref{eqn:conditional_u_cov} for $U$-$T$ relationship, the confounding bias is bounded by
\begin{equation}
\label{eqn:hetero_bound2}
    |\text{Bias}_{a,t_1,t_2}| 
    \leq \frac{1}{\sigma} \sqrt{\frac{R_{T \sim U \mid X}^2}{1 - R_{T \sim U \mid X}^2}} E_X\left[\parallel a' \Gamma_{t_1,x}(t_1 - \mu_{t\mid x})\parallel_2 + \parallel a' \Gamma_{t_2,x}(t_2 - \mu_{t\mid x})\parallel_2 \right].
\end{equation}
Under factor confounding, the only unidentified parameters are $\rho_x$ and $\rsqtx$ respectively. 

\begin{proof}
Under the above assumptions, the confounding bias is equal to
\begin{align}
    \text{Bias}_{a,t_1,t_2} &= (E[a'Y \mid t_1] - E[a'Y \mid t_2]) - (E[a'Y \mid do(t_1)] - E[a'Y \mid do(t_2)]) \\
    &= E_X[a'(\Gamma_{t_1,x}\Sigma_{u\mid t, x}^{-1/2}\mu_{u\mid t_1,x} - \Gamma_{t_2,x}\Sigma_{u\mid t, x}^{-1/2}\mu_{u\mid t_2,x})] \\
    &= E_X\left[\frac{a'\Gamma_{t_1,x}\beta_x}{\sigma_{t\mid x}\sqrt{\sigma^2_{t\mid x} - \beta_x^{\prime}\beta_x}}(t_1 - \mu_{t\mid x}) - \frac{a'\Gamma_{t_2,x}\beta_x}{\sigma_{t\mid x}\sqrt{\sigma^2_{t\mid x} - \beta_x^{\prime}\beta_x}}(t_2 - \mu_{t\mid x})\right].
 \end{align}
By Cauchy–Schwarz inequality, we have 
\begin{equation}
\left| \frac{a'\Gamma_{t_i,x}\beta_x(t_i - \mu_{t\mid x})}{\sigma_{t\mid x}\sqrt{\sigma^2_{t\mid x} - \beta_x^{\prime}\beta_x}}\right| \leq \frac{1}{\sigma_{t\mid x}} \sqrt{\frac{\rho_x^2}{1 - \rho_x^2}}\parallel a' \Gamma_{t_i,x}(t_i - \mu_{t\mid x})\parallel_2,
\end{equation}
for $i \in \{1,2\}$. Then the magnitude of confounding bias is bounded by 
\begin{align}
|\text{Bias}_{a,t_1,t_2}| &\leq \left|E_X\left[ \frac{a'\Gamma_{t_1,x}\beta_x(t_1 - \mu_{t\mid x})}{\sigma_{t\mid x}\sqrt{\sigma^2_{t\mid x} - \beta_x^{\prime}\beta_x}} \right]\right| + \left|E_X\left[ \frac{a'\Gamma_{t_2,x}\beta_x(t_2 - \mu_{t\mid x})}{\sigma_{t\mid x}\sqrt{\sigma^2_{t\mid x} - \beta_x^{\prime}\beta_x}} \right]\right| \\
&\leq E_X\left[\frac{1}{\sigma_{t\mid x}} \sqrt{\frac{\rho_x^2}{1 - \rho_x^2}}(\parallel a' \Gamma_{t_1,x}(t_1 - \mu_{t\mid x})\parallel_2 + \parallel a' \Gamma_{t_2,x}(t_2 - \mu_{t\mid x})\parallel_2) \right].
\end{align}
The bound is achieved when $\beta_x$, $\Gamma_{t_1,x}'a$ and $\Gamma_{t_2,x}'a$ are collinear, $a'\Gamma_{t_1,x}\beta_x(t_1 - \mu_{t\mid x})$ and $a'\Gamma_{t_2,x}\beta_x(t_2 - \mu_{t\mid x})$ have constant opposite signs for all $x$. Under parameterization \eqref{eqn:conditional_u_mean}-\eqref{eqn:conditional_u_cov}, we obtain inequality \eqref{eqn:hetero_bound2} by moving constants out of the expectation.
\end{proof}
\end{theorem}

\pagebreak

\section{Additional Results}
\label{sec:alcohol_additional}
To further understand the plausibility of the factor confounding assumption, we examine the implied values of $R^2_{a'Y\sim U | T, X}$, the partial variance in each outcome that is explained by the latent factors under factor confounding (see Appendix \ref{sec:cal_r2y}).  In Table \ref{tab:obs_partial_rsq} we report $R^2_{a'Y\sim U | T, X}$ for each outcome under factor confounding and compare these values to the partial fraction of variance explained by each of gender, age, and education. We find that while the observed covariates are significantly associated with the outcomes, they explain a small fraction of the total variation in them (mostly less than $0.1$).  In contrast, we find that for the vast majority of outcomes, latent factors explain a relatively large fraction of the outcome variance, with $\rsqytx$ greater than $0.3$ for most outcomes.  For these outcomes, we might expect that the bounds computed under factor confounding are conservative.  Two important exceptions are methylmercury and potassium for which the values for $R^2_{a'Y\sim T, U | X}$ are  $0.07$ and $0.05$ respectively, the smallest values across all outcomes.  The inferred latent factors seem to explain far less residual variation in methylmercury and potassium than they do for most other measured outcomes. 

\begin{figure}	
\centering
\includegraphics[width=0.8\textwidth]{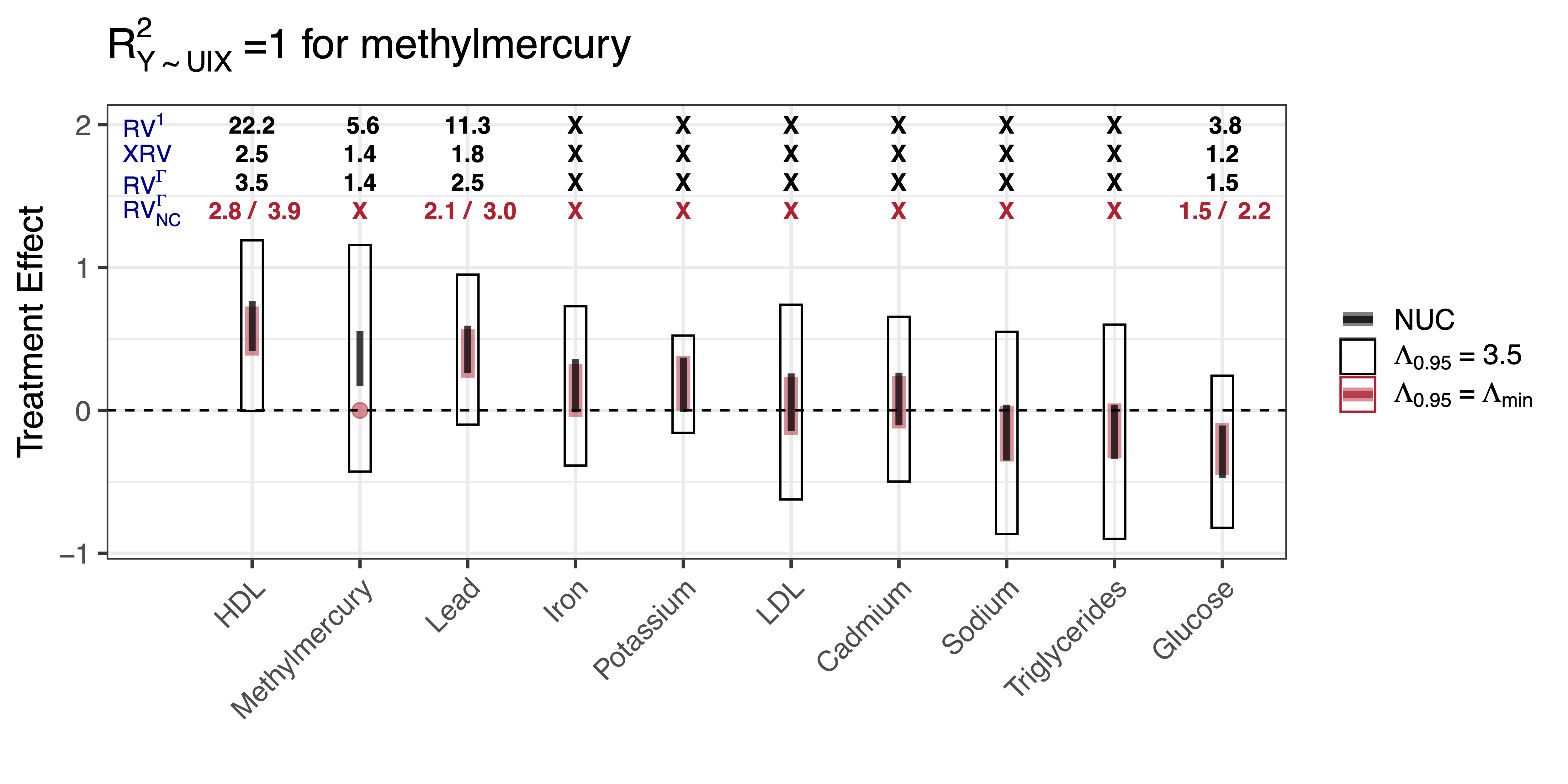}
    \caption{ 95\% posterior credible intervals for the causal effects of light drinking on each outcome under NUC (black) and	under factor confounding for all outcomes except for methylmercury, for which we assume  $R^2_{a'Y \sim U | T, X} = 1$, so that all residual outcome variance in the null control is due to confounders (predominantly single outcome confounding). $\Lambda_{0.95}=3.5$ matches the magnitude of the multiplicative change in odds of light drinking when adding age into a propensity model which includes gender and education. In red, we plot the 95\% credible intervals when methylmercury is assumed to be a null control outcome and $\Lambda_{0.95} = \Lambda_{min, \mathcal{C}}$, the value needed to nullify methylmercury under factor confounding (red intervals).  Numbers in black indicate different robustness values in the $\Lambda_{0.95}$-parameterization, including the single-outcome robustness ($RV^1$), extreme robustness ($XRV$), robustness under factor confounding ($RV^\Gamma$) and the combined robustness with methylmercury as a null control, $RV^\Gamma_{c=2} \geq \Lambda_{min, \mathcal{C}}$ (in red).  In this setting, the value needed to nullify mercury drops to $\Lambda_{min, \mathcal{C}}=1.4$.  The null control assumption is considerably weaker in this case, and thus has less influence on non-null outcomes.	 Listed values are conservatively reported as the lower endpoints of the 95\% posterior credible intervals of the robustness values. \label{fig:interval_plots2}}
\end{figure}

The small value for mercury might indicate that there are environmental confounders that influence both mercury levels and the propensity for light drinking, but do not influence the other outcomes, thus violating factor confounding for methylmercury.  To account for this possibility, we first establish robustness under the most conservative assumption about the outcome-confounder dependence for mercury, by fixing $a'\Gamma=a'\Sigma_{y\mid t, x}^{1/2}$ so that $R^2_{a'Y\sim U|T, X}=1$.  In Figure \ref{fig:interval_plots2}, we plot the posterior intervals and report robustness values of $\Lambda_{0.95}$ when all the variance in this null control outcome is due to confounders. When $R^2_{a'Y\sim U|T, X}=1$ for mercury, the strength of treatment-confounder association needed to satisfy the null control assumption drops from $\Lambda_{0.95} = 4.2$ to $\Lambda_{0.95} = 1.4$. Additionally, the bias correction for non-null outcomes is very small when fixing $R^2_{a'Y\sim U|T, X}=1$ for methylmercury, since $1-0.07 = 0.93$ of the residual variance in methylmercury is driven by confounders which are uncorrelated with all other outcomes. In this case, the posterior interval for potassium changes very little and still overlaps zero. Relative to NUC, the qualitative conclusions for all outcomes remain essentially unchanged, although the robustness for each significant outcome increases slightly.   

These results are consistent with the qualitative conclusions from \citet{rosenbaum2021sensitivity}, who argues that the apparent presence of confounding bias for methylmercury strengthens the robustness of the finding that light drinking increases HDL cholesterol.  Likewise, we show that when $\rsqytx=1$ for methylmercury, the robustness of HDL increases from $\Lambda^{RV}_{0.95} = 3.5$ to $\Lambda^{RV}_{0.95, \mathcal{C}} = 3.9$ under the null control assumption.  

% \begin{figure}[h!]
%     \centering
%     \includegraphics[width = 0.5\textwidth]{figs/R2_scale_function.pdf}
%     \caption{Confounding bias as a function of $R_{T \sim U}^2$. The bias is proportional to $\sqrt{R^2 / (1 - R^2)}$ which is approximately linear until $R^2$ is larger than about 0.95, at which point it explodes.}
%     \label{fig:R2_scale_function}
% \end{figure}

% \begin{table}[h!]
% \centering
% \begin{tabular}{l|r|r|}
%   & \textbf{ELPD Difference} & \textbf{SE}\\
% \hline
% Rank 4 & 0.0 & 0.0\\
% \hline
% Rank 6 & -0.5 & 3.3\\
% \hline
% Rank 5 & -1.2 & 1.8\\
% \hline
% Rank 3 & -12.8 & 5.7\\
% \hline
% Rank 2  & -30.0 & 9.4\\
% \hline
% Rank 1 & -63.1 & 13.6\\
% \hline
% \end{tabular}
% \caption{Difference expected log posterior density (ELPD) for ranks 1 through 6 and standard error.  Results were computed using \texttt{loo} package \citep{loo}.  Ranks 4-6 all have ELPD which are within two standard errors of each other, and thus we cannot reliably distinguish the model fit. \label{tab:elpd}}
% \end{table}

% latex table generated in R 4.1.2 by xtable 1.8-4 package
% Fri May 27 10:14:38 2022
\begin{table}[ht]
\centering
\begin{tabular}{l|c|c|c|c|}
 & $R^2_{Y \sim Age | -Age}$ & $R^2_{Y \sim Gender | -Gender}$ & $R^2_{Y \sim Education | -Education}$ & $R^2_{Y\sim U | T, X}$ (fact. conf.) \\ 
\hline 
HDL & 0.013 & 0.12 & 0.01 & 0.483 \\ 
  Methylmercury & 0.019 & NS & 0.02 & 0.067 \\ 
  Glucose & 0.092 & 0.014 & 0.005 & 0.181 \\ 
  Potassium & 0.005 & 0.045 & NS & 0.049 \\ 
  Sodium & 0.008 & 0.003 & NS & 0.635 \\ 
  Iron & 0.007 & 0.071 & 0.006 & 0.34 \\ 
  Lead & 0.215 & 0.047 & 0.014 & 0.376 \\ 
  LDL & NS & NS & NS & 0.511 \\ 
  Triglycerides & 0.034 & 0.009 & 0.003 & 0.862 \\ 
  Cadmium & 0.041 & 0.028 & 0.013 & 0.345 \\ 
   \hline
\end{tabular}
\caption{Partial coefficient of determination, $R^2_{Y \sim X_i | T, X_{-i}}$ for each outcome for age, gender, and education (first 3 columns) and the partial coefficient of determination for unobserved confounders given observed confounders under the $\Gamma$-confounding assumption.  ``NS'' the covariate was not a significant predictor for the outcome. $R^2_{Y \sim U \mid T, X}$ under $\Gamma$-confounding is significantly larger than the observable counterparts, although it is by far the smallest for methylmercury and potassium.  It may be important to consider the possibility of single outcome confounding for these outcomes.   \label{tab:obs_partial_rsq}}
\end{table}

\begin{table}[h!]
\centering
\begin{tabular}{l|r|r|}
 & ELPD Difference & Standard Error of Difference \\ 
  \hline
Full rank & 0.00 & 0.00 \\ 
\hline
Rank 6 & -5.13 & 16.61 \\ 
\hline
Rank 5 & -7.00 & 15.53 \\ 
\hline
Rank 4 & -21.70 & 17.91 \\ 
\hline
Rank 3 & -48.17 & 19.70 \\ 
\hline
Rank 2 & -102.69 & 17.01 \\ 
\hline
Rank 1 & -223.20 & 24.33 \\ 
   \hline
\end{tabular}
\caption{Differences in expected log posterior density (ELPD) for ranks 1 through 6 and for full rank, with associated standard errors.  Results are computed using \texttt{loo} package \citep{loo}. Ranks 4-6 all have an ELPD which is within two standard errors of the full rank model, which suggests that these are reasonable models for the observed data. Models with ranks 1-3 are insufficient to capture correlations between the outcomes. \label{tab:elpd}}
\end{table}

\begin{figure}[h!]
    \centering
    \includegraphics[width = 0.7\textwidth]{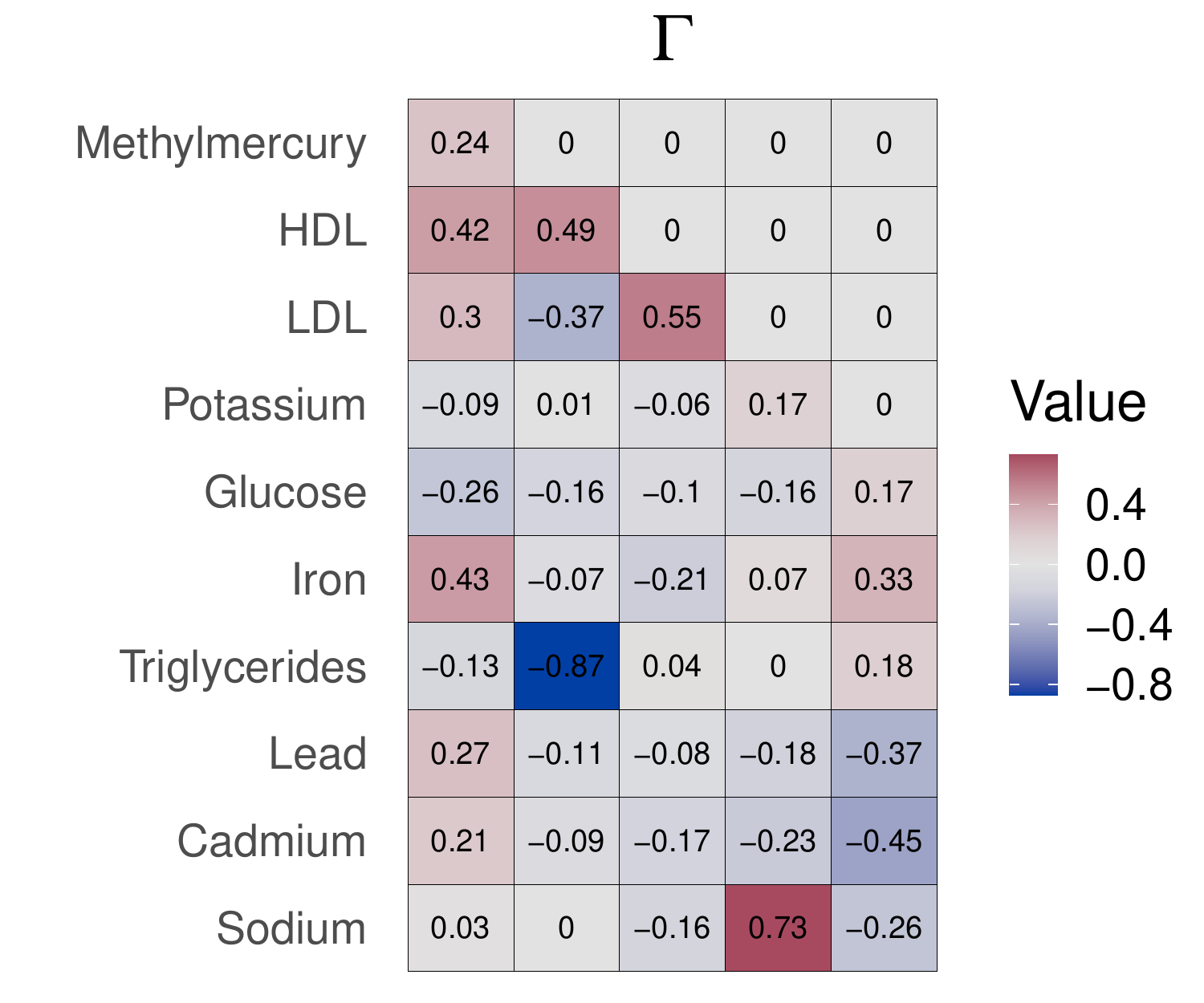}
    \caption{Heatmap of the $\hat \Gamma$ from the NHANES example under $\Gamma$-confounding. $\Gamma$ is only identifiable up to rotations from the right, so for visualization purposes, we choose a rotation which makes the upper triangle of the first four rows zero. The first column thus indicates the direction and magnitude of the change in the partial identification regions for each outcome when methylmercury is an assumed null control.  If we were to fix $R^2_{a'Y\sim U | T,X} = 1$ for methylmercury then we would simply change the first cell of the first row to $3.24$}
    \label{fig:nhanes_heat}
\end{figure}

\end{document}